\documentclass{appolb}
\usepackage{graphicx}
\usepackage[utf8]{inputenc}
\usepackage{amsmath}
\usepackage[english]{babel}
\usepackage{graphicx}
\usepackage{multicol}
\usepackage{subfigure}
\usepackage{algorithm,algorithmic}
\usepackage[numbers,sort&compress]{natbib}
\usepackage{color}
\usepackage{epstopdf} 
\usepackage{multirow}
\usepackage{amssymb}
\begin{document}
% \eqsec  % uncomment this line to get equations numbered by (sec.num)
\title{Hierarchical community structure in complex (social) networks.%
\thanks{Presented at SummerSoltice 2013}%
% you can use '\\' to break lines
}
\author{Emanuele Massaro$^{\S,\dagger}$, Franco Bagnoli$^\P$
\address{$^\S$Dept. of Information Engineering and CSDC, University of Florence, Italy\\
$^\dagger$Now at Dept. of Civil and Environmental Engineering\\ Carnegie Mellon University, Pittsburgh, PA 15213-3890, USA }
\address{$^\P$Dept. of Physics and Astronomy, and CSDC, University of Florence, \\also INFN sez. Firenze, Italy}
}
\maketitle
The investigation of community structures in networks is a task of great importance in many disciplines, namely physics, sociology, biology and computer science, where systems are often represented as graphs. One of the challenges is to find local communities in a graph from a local viewpoint,  in the absence of the access to global information,  and to reproduce the subjective hierarchical vision for each vertex. In this paper we present the improvement of an information dynamics algorithm in which the label propagation of nodes is based on the Markovian flow of information in the network under cognitive-inspired constraints~\cite{Massaro2012}. We introduced two more complex heuristics that allow to detect the hierarchical community structure of the networks from a source vertex or a community, adopting fixed values of model's parameters. Experimental results show that the proposed methods are efficient and well-behaved in both real-world and synthetic networks.

\PACS{PACS 07.05.Kf, 89.65.-s, 89.75.Fb}
  
\section{Introduction}
We live in a world of networks. The networks around us and ourselves, as people, we are part of the network of social relations between individuals. Examples of networks in the world are the WWW, the rail network, the subway, neural networks, the telephone network, or less concrete entities, such as the relations of knowledge and collaboration between people. In general, a graph or network is a very general approximation of a system constituted by many entities, called nodes (persons, computers, proteins, chemicals, etc.) linked to each other and interacting through connections (which may be, therefore, cables among computers, hyperlinks between web pages, collaborations among people,  reactions among chemical substances, etc..). The study of networks is therefore very relevant in many disciplines, given the wide variety of structures and systems of the real world that can be incorporated into the category of ``complex networks''. 

A complex network is a system with non trivial topological features which would  not be detectable in simple graphs. The majority of real networks, being them social, biological or technological, may be considered complex: this depends on some features such as, \textit{e.g.}, the nodes' degree of distribution~\cite{Erdos, Barabasi99}, the assortativity among  vertexes~\cite{Newman2002}, relatively  short path lengths~\cite{albert1999, Watts1998} and, often,  an evident hierarchical structure~\cite{ravasz2003}.  
\begin{figure}[t]
\begin{center}
\includegraphics[width=.5\columnwidth]{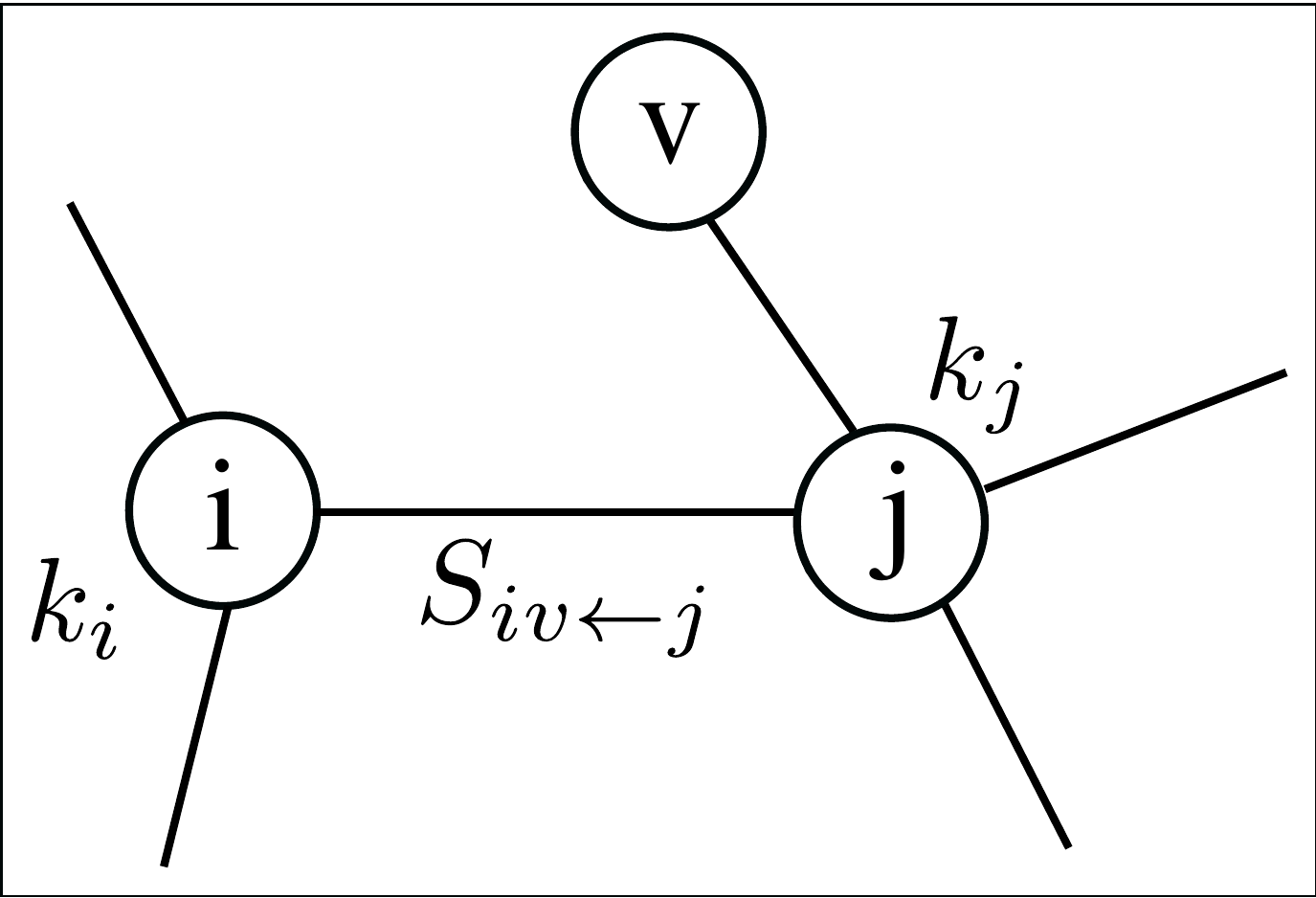} 
\end{center}
\caption{
Communication phase. Let us consider the node $i$ that communicates with the node $j$. The node $j$  passes his knowledge of the world to the node $i$. Here we can consider three different scenarios: the first one takes into account that the knowledge depends on the connection degree of the incoming node. In this way the most connected nodes have less time to talk with others. In the second  hypothesis the spread information depends of the connectivity degree of the selected nodes: then  Eq.~\eqref{information} becomes $S_{iv \leftarrow j} = mS_{iv} + (1-m){S_{jk}}/{k_i}$. In the third one it depends both on the degree of node $i$ and on the degree of node $j$: $S_{iv \leftarrow j} = mS_{iv} + (1-m){S_{jv}}/{\sqrt{k_i^2 + k_j^2}}$.}
\label{fig:ij}
\end{figure}
Another important  feature of complex networks is the presence of a community structure~\cite{Girvan02}. Many real networks  are not homogeneous and do not consist of a single block of indistinct nodes, rather they exhibit community structures, \textit{i.e.},  groups of vertexes exhibiting a high densities of internal arches, compared to a relatively lower number of connections with other groups.

In recent years, the possibility of automatically identifying communities within networks  has been explored in detail,  giving rise to a new field of research called ``community detection''~\cite{Fortunato, Newman:2004fk}. The problem of community detection has attracted the attention of various researchers coming from different disciplines from physics to social sciences. Nowadays, no common definition for community has been agreed upon: it is widely accepted that a community is a group of vertexes more linked than between the group and the rest of the graph. 

This is clearly a poor definition, and indeed, on a connected graph, there is not a clear distinction between a community and the rest of the graph. In general, there is a continuum of nested communities whose boundaries are somewhat arbitrary. Moreover, in complex networks, and in particular in social networks, it is very difficult to give a clear definition of the community concept: this is due to the fact that nodes are often attributed to overlapping communities,  since they belong to more than one cluster or module or community. For instance, people usually belong to different communities at the same time, depending on their families, friends, colleagues, etc.~\cite{Palla}: so people, making a subjective screening of the network in which they live, have their local vision of communities to which they belong. 

In order to have a subjective vision of the surrounding world it is necessary to develop community detection algorithms based on the local link structure of the network, algorithms that work by getting information from neighbouring sites~\cite{Bagrow2008, Bagrow20081, Clauset2005, Luo2008, Lanc09, Huang11}. 

In this paper we show a method for detecting communities  where each node discovers its representation of the network by ``talking'' with the neighbors and then elaborating the information trough a cognitive-inspired mechanism~\cite{Massaro2012}. In particular we introduce more accurate heuristics in order to free the method from the tuning of parameters. The main feature of these two heuristics called respectively \emph{IDA + LTE} and \emph{Double pruning} is to reveal  the hierarchical community structure of complex networks from a local viewpoint. We demonstrate that a node can discover the network's multi-levels giving its subjective vision of the world until it wants or can explore the graph without needing any global information about it.  

The reminder of the paper is organized as follows. First, in Section~\ref{method}, the general method~\cite{Massaro2012} is presented. Then we introduce two more complex heuristics applied to this model in order to reveal the multi-resolution community structure of both synthetic and real networks. In the results we also compare out method with other well-known community detection algorithm by evaluating the so-called Normalized Mutual Information (NMI) on LFR benchmarks~\cite{LFR}. The paper is closed by final remarks and conclusions.
 
\begin{figure}[t!]
\centering
{\includegraphics[width=.5\columnwidth]{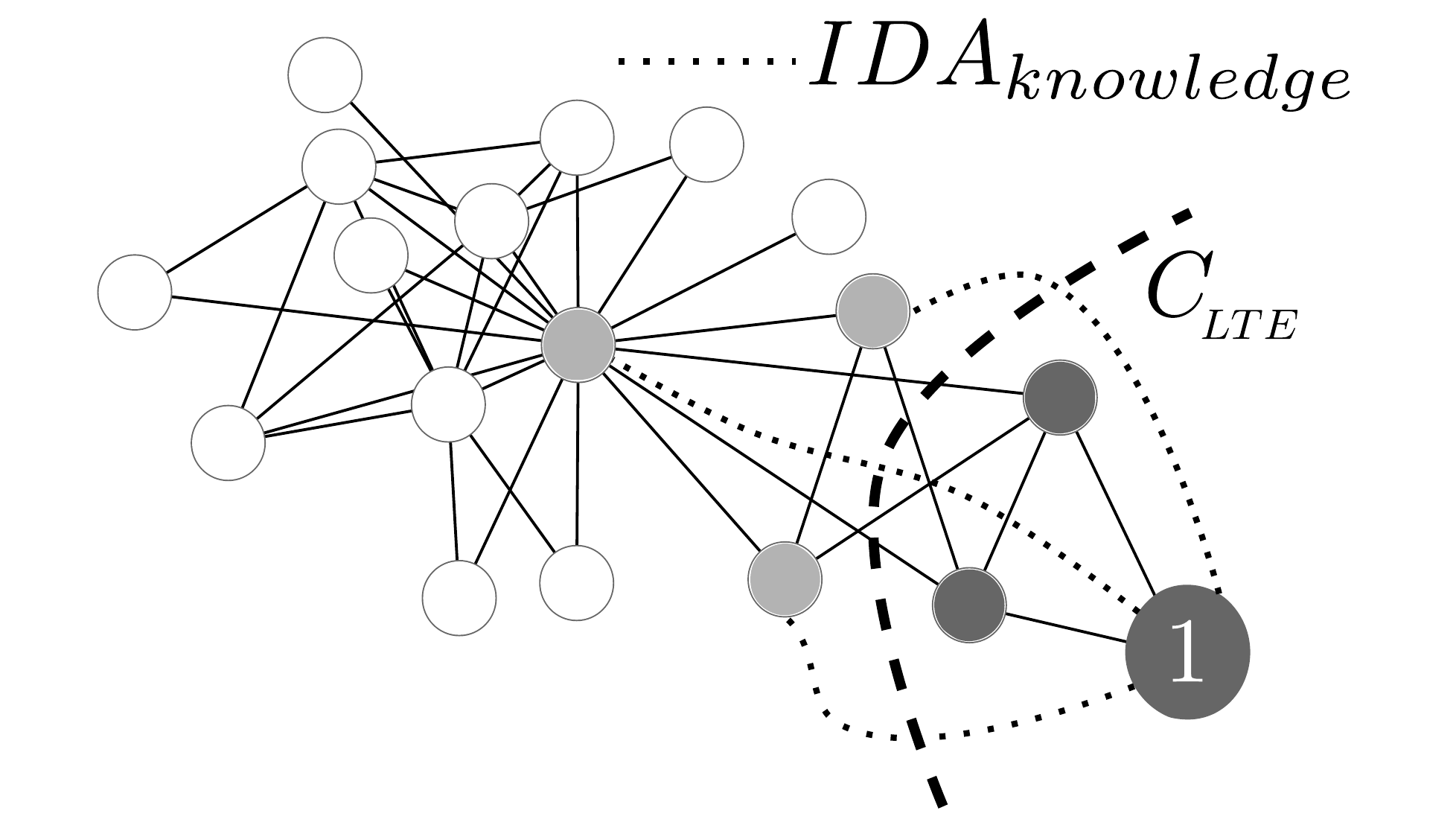}}
\caption{\label{fig:nuovoalgortimschema} Schematic rapresentation of IDA + LTE algorithm. We suppose that $LTE$ detect for the node 1 the dark-gray nodes as community nodes ($C_{LTE}$): after that we compute the $IDA$ on the selected nodes. The node takes the information from those accepted as its community creating a virtual link with these (dotted line in figure). The $LTE$ decides whether to accept the new nodes memebrs of own community or not.}
\end{figure}

\section{The Method}
\label{method}
We consider $N$ individuals, labeled from $1$ to $N$. Let us denote by $A$ the adjacency matrix, $A_{ij} = 1 $ (0) indicates the presence (absence) of a link from site $j$ to site $i$. Each individual $i$ is characterized by a state vector $S_i$, representing his/her knowledge of the outer world. 
We interpret $S$ as a probability distribution, assuming that $S_i^{(v)}$ is the probability that individual $i$ belongs to the community $v$. Thus, $S_i^{(v)}$ is normalized on the index $v$. We shall denote with $S=S(t)$ the state of the all network at time $t$, with $S_{iv} = S_i^{(v)}$. We shall initialize the system by setting $S_{ij}(0) = \delta_{ij}$, where $\delta$ is the Kroneker delta, $\delta_{ij}=1$ if $i=j$ and zero otherwise. 

In other words, at time 0 each node knows only about itself.  In this implementation we assume that each individual spends the same amount of time in communications, so the information delivered depends on connection degree. Let us consider the node $i$ that communicates with the node $j$ as shown in  Figure~\ref{fig:ij}. We can assume that people with more connections dedicate less time to each of them. Since the amount of available time is limited we say that the knowledge of  node $i$ about node $j$ depends on the connectivity degree of $j$, then $S_{i \leftarrow j}$ is function of the connectivity degree $k$ of node $j$. Alternatively, we can consider  that the information depends on the connectivity of $i$, in this way we give much more importance to the \emph{speaking phase}.

\begin{figure}[t!]
\renewcommand*\thesubfigure{\arabic{subfigure}}
\centering
\subfigure[] 
{\includegraphics[width=4cm]{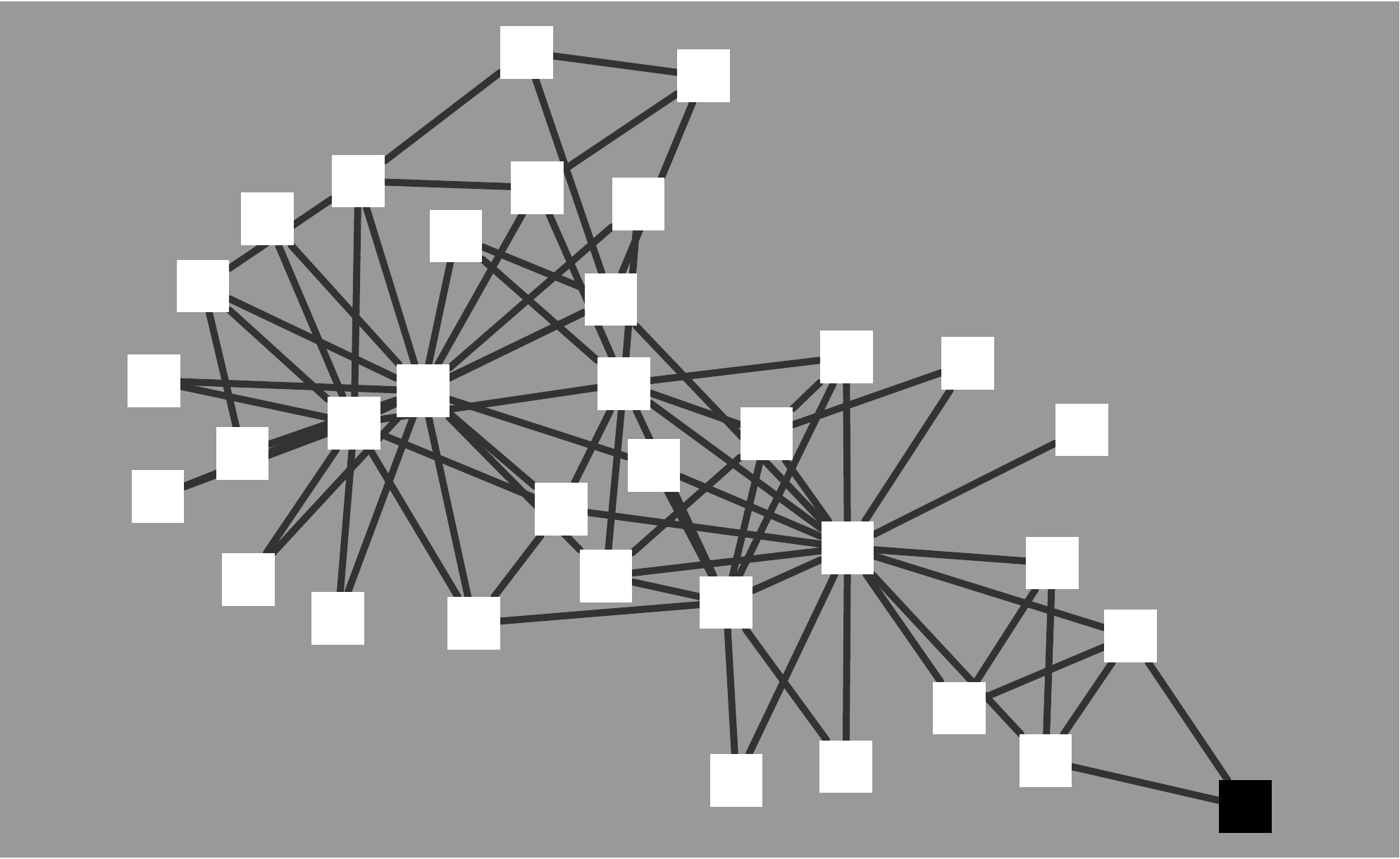}}
\hspace{0mm}
\subfigure[]
{\includegraphics[width=4cm]{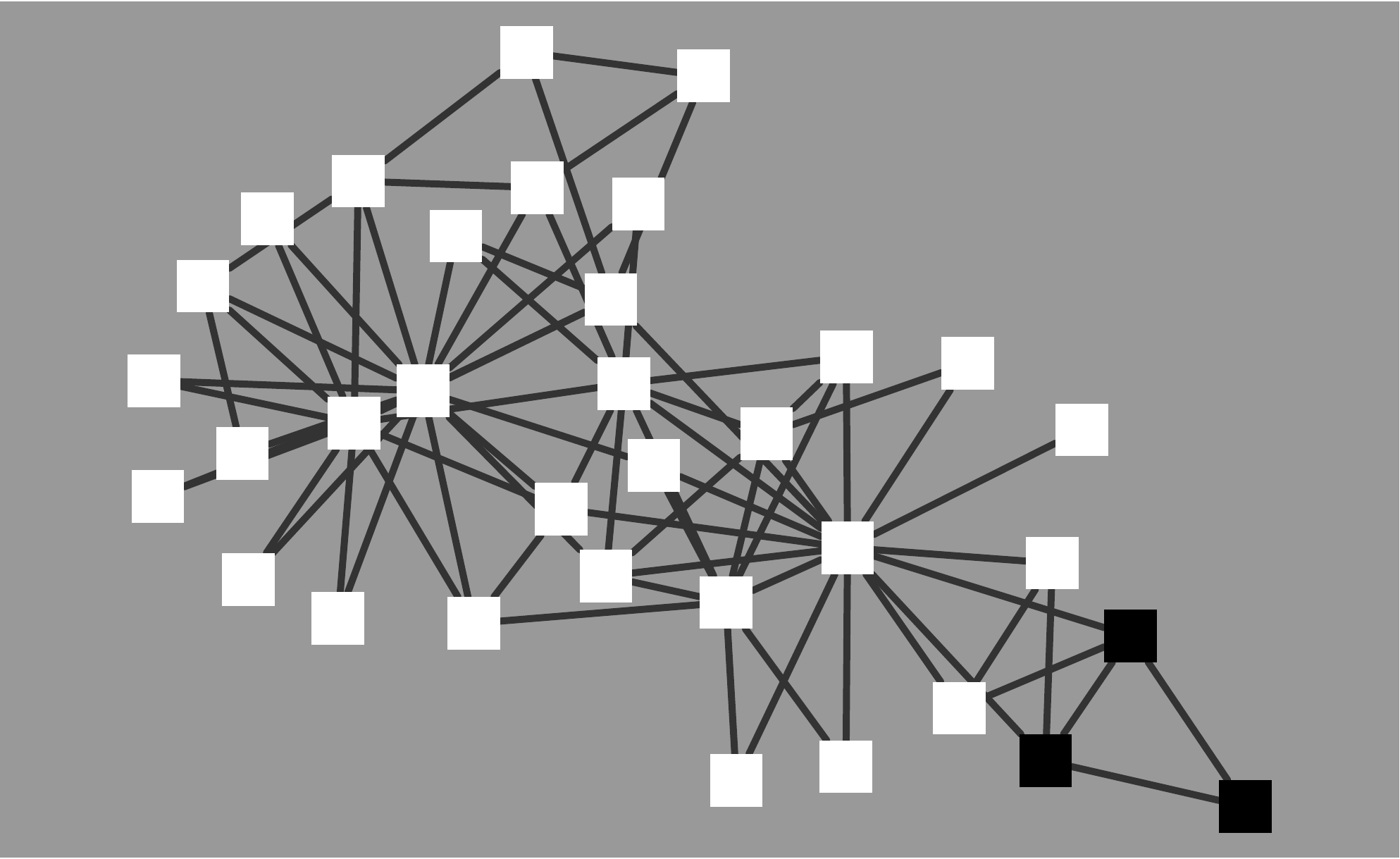}}
\hspace{0mm}
\subfigure[]
{\includegraphics[width=4cm]{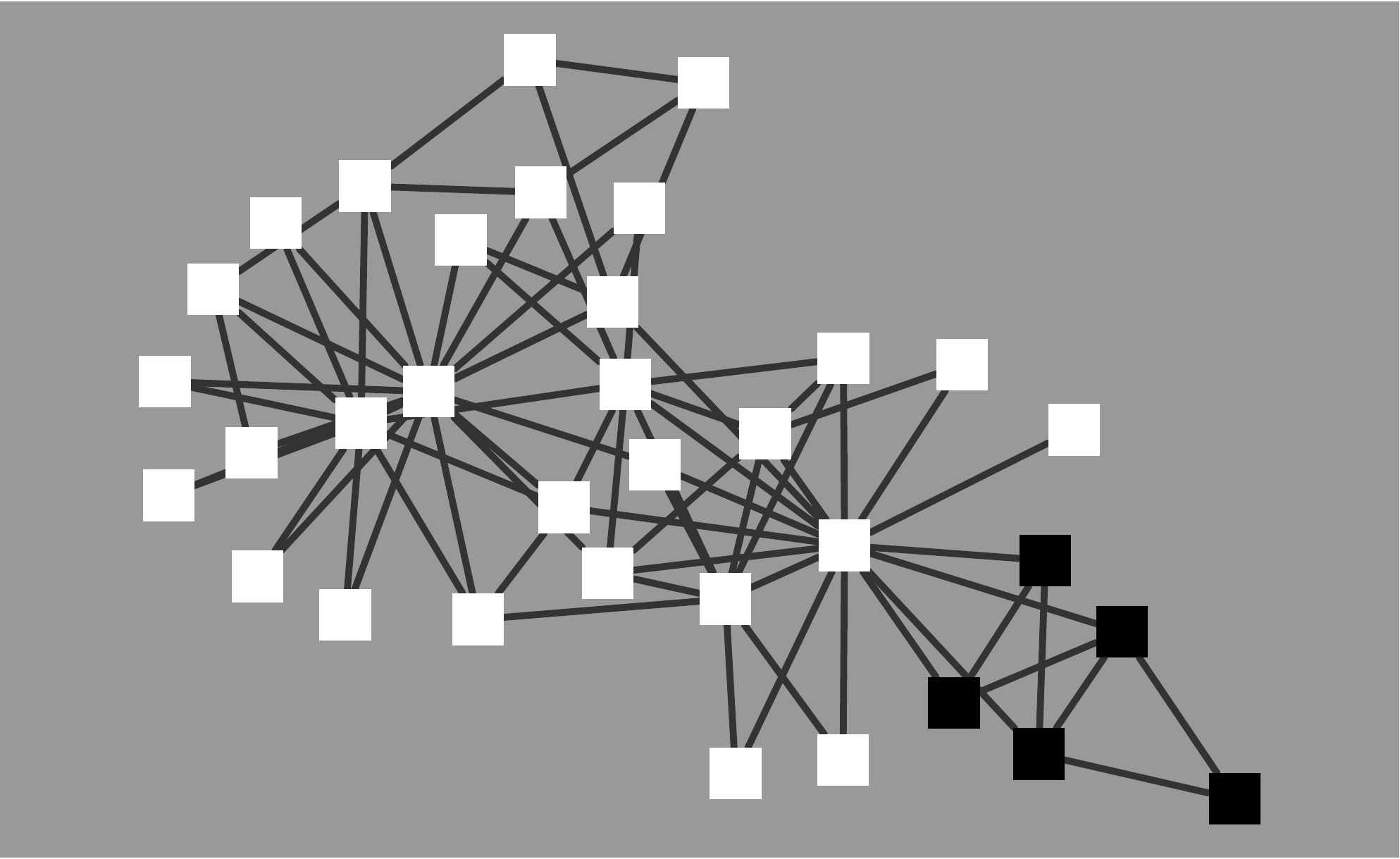}}
\hspace{0mm}
\subfigure[]
{\includegraphics[width=4cm]{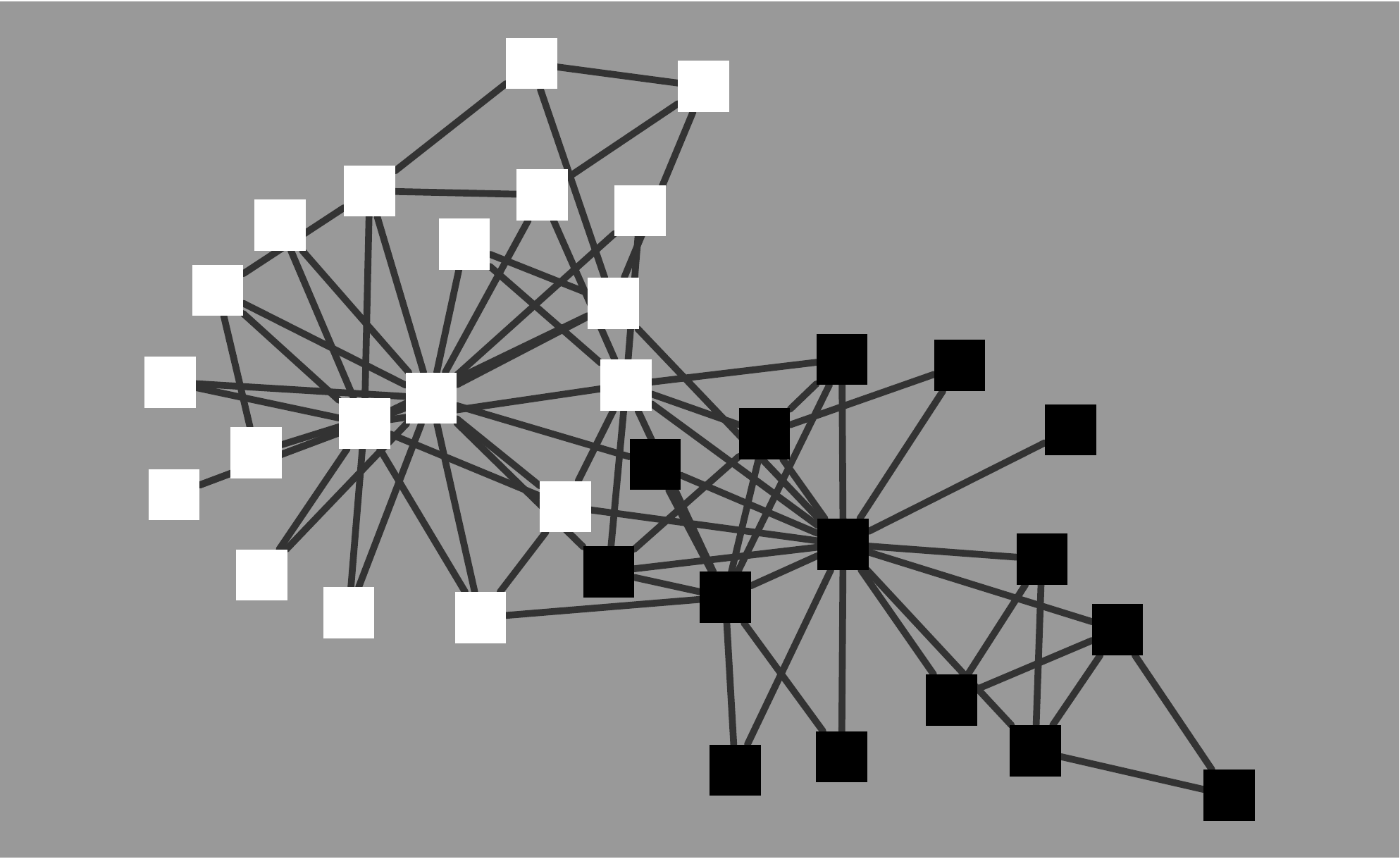}}
\hspace{0mm}
\subfigure[]
{\includegraphics[width=4cm]{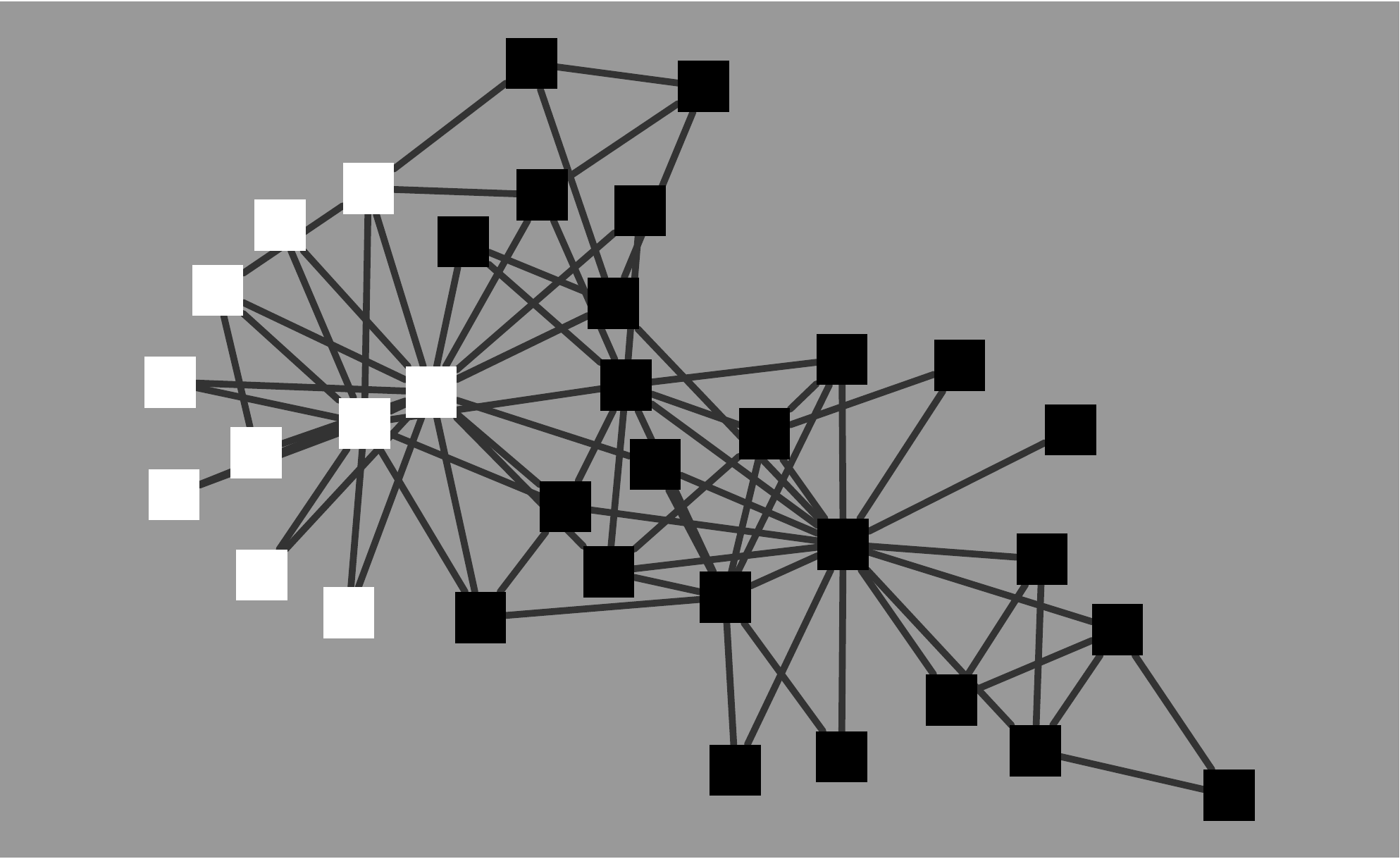}}
\hspace{0mm}
\subfigure[]
{\includegraphics[width=4cm]{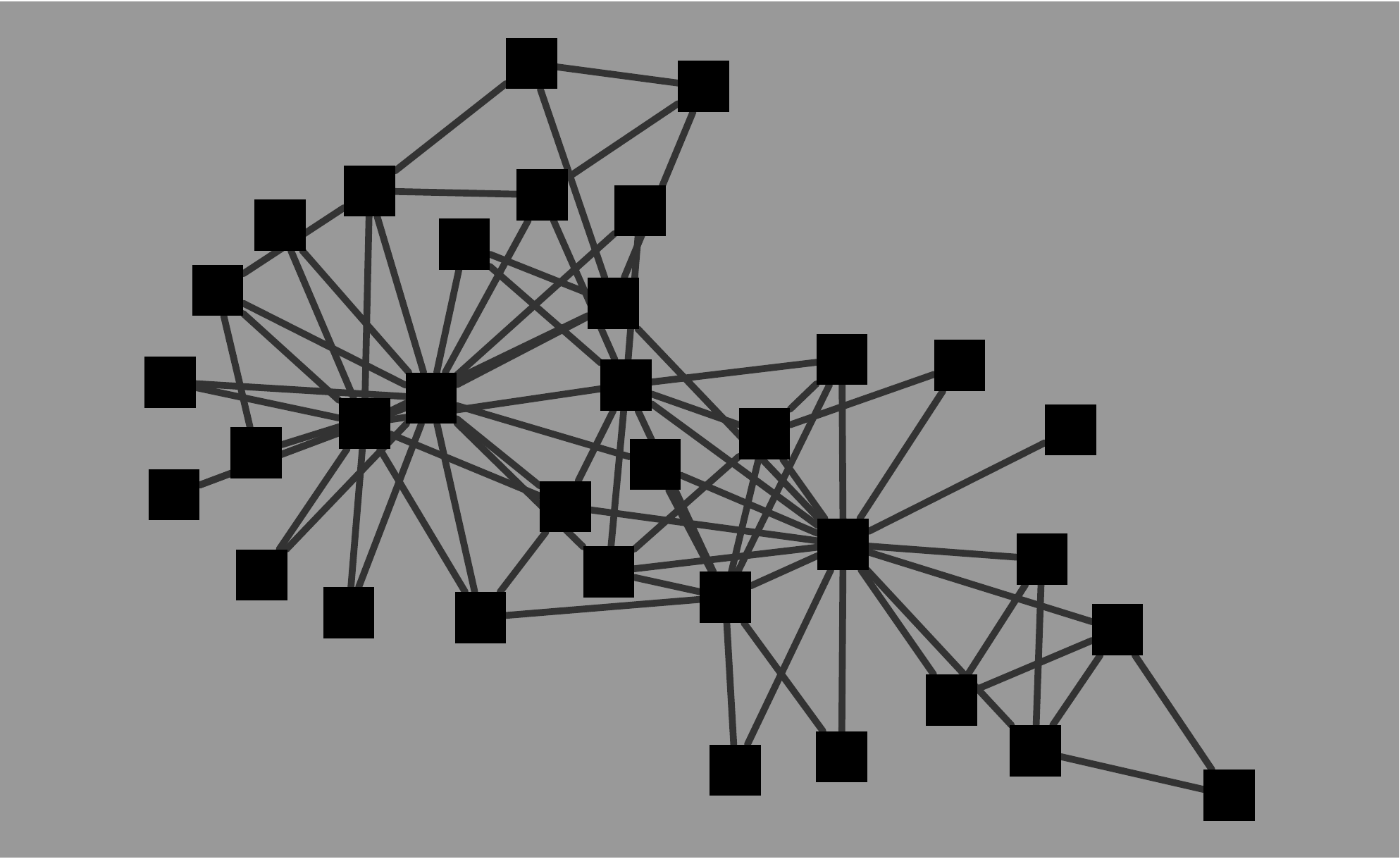}}
\caption{\label{fig:karate_completa} Node $17$ in the \emph{Zachary's Karate Club} network. In the first step the LTE algorithm accept 3 nodes for the local community. After that we compute the IDA + LTE algorithm: the knowledge of the other nodes permits to discover the structure of the whole network. From the starting vertex (black node at $T=1$) at time $T=2$ the community nodes are 17-6-7 (black nodes)and finally all the network is discovered by the node 17 thanks the information dynamics algorithm.}
\end{figure} 

\begin{figure}[t]
\begin{center}
\includegraphics[width=.6\columnwidth]{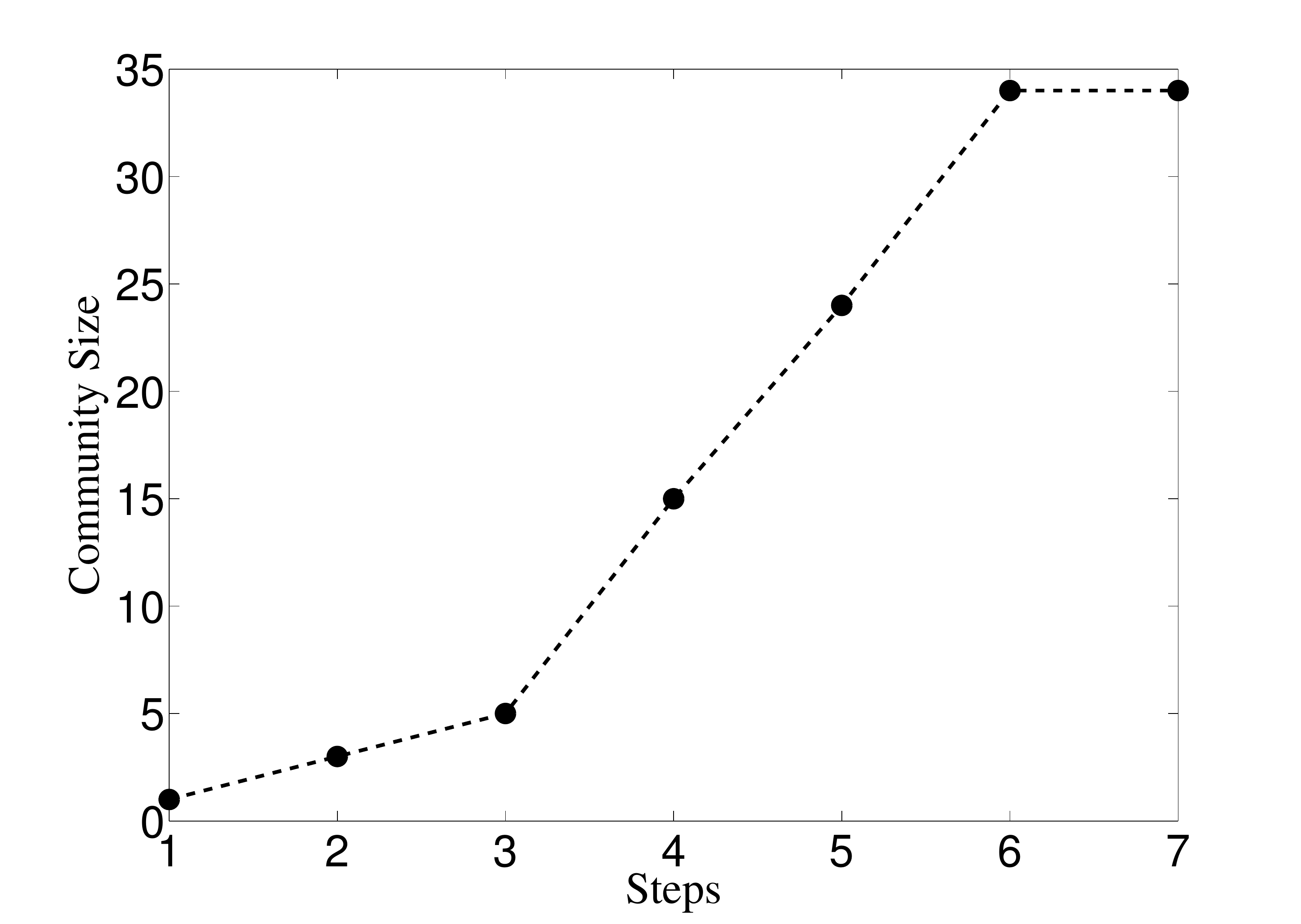} 
\end{center}
\caption{Temporal evolution (\emph{x-axis}) of community size ($y-axis$) for the node 17 in the \emph{Zachary's Karate Club} network. }
\label{fig:karate_temp}
\end{figure}

The dynamics of the information is given by an alternation of communication and elaboration phases. 
Communication is implemented as a simple diffusion process, with memory $m$. The memory parameter $m$ allows us to introduce some limitations in human cognitive such as the mechanism of oblivion and the timing effects: in fact the most recent information has more relevance than the information gathered in the past.  The knowledge of a generic node $v$ for the node $i$ trough $j$ occurs  via a communication phase where the state of the system evolves as
\begin{equation}
S_{iv \leftarrow j} = mS_{iv} + (1-m)\frac{S_{jv}}{k_j}.
\label{information}
\end{equation}

The representation of the model is consistent with the ecological rationality of the system since the knowledge is distributed among the network elements (state vectors $S_i$). Our model is an improvement  of the MCL algorithm proposed by Van Dongen~\cite{MCL} that simulates a random walk within the graph by an alternation of two phases called expansion and inflation. In our method we consider the Markov matrix $M_{i,j} = \frac{A_{ij}}{k_i}$, where $k_i$ is the connectivity degree of node $i$. Finally, the evolution of the system is given by the sequence $S^0 \rightarrow S^1 \rightarrow S^2$.

The communication phase based on the connectivity of incoming nodes, is given by the following:
\begin{equation}
S^1 = mS^0 + (1-m)M'S^0,
\end{equation} 
where $0 < m < 1$ is a memory parameter and $M'$ is the  transpose of $M$ . In this way the resulting matrix is still stochastic,\textit{ i.e.}, the matrix elements (on each rows) correspond to probability values.
The second phase, called elaboration or inflation is implemented as the rising of each element of the the matrix $S_1$ to the power $\alpha$,
where $\alpha > 1$ corresponds to the inflation parameter:
\begin{equation}
S^2_{ij}=\left(S^{1}_{ij}\right)^{\alpha}
\end{equation}

After this phase,  in order to have a probability matrix we normalize the matrix $S_2$  over the columns. We assume that individuals have a large enough memory so that they can keep track of all information about all other individuals. In a real case, one should limit this memory and apply an input/output filtering.

\begin{figure}[t!]
\renewcommand*\thesubfigure{\arabic{subfigure}}
\centering
\subfigure[] 
{\includegraphics[width=4cm]{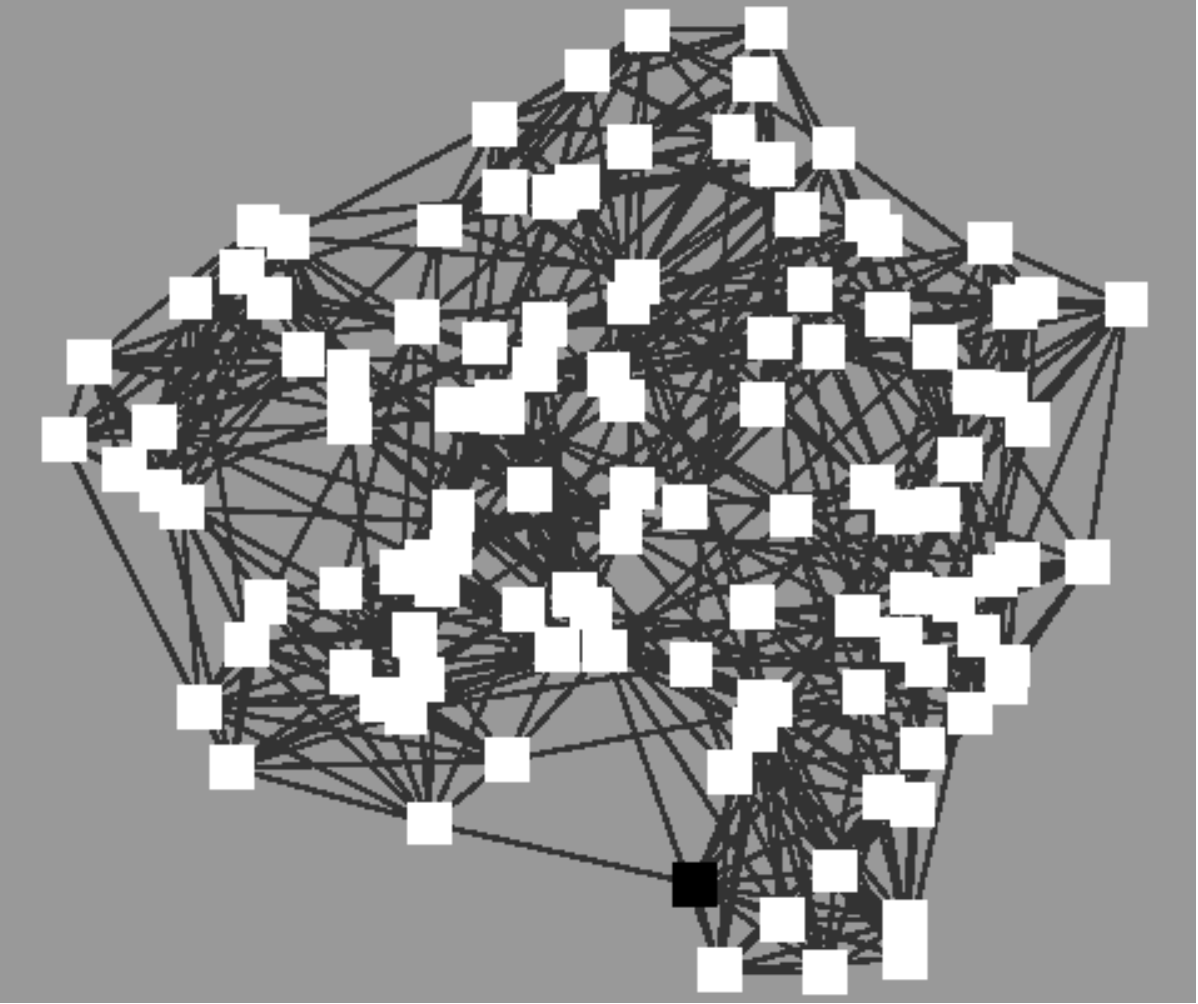}}
\hspace{0mm}
\subfigure[]
{\includegraphics[width=4cm]{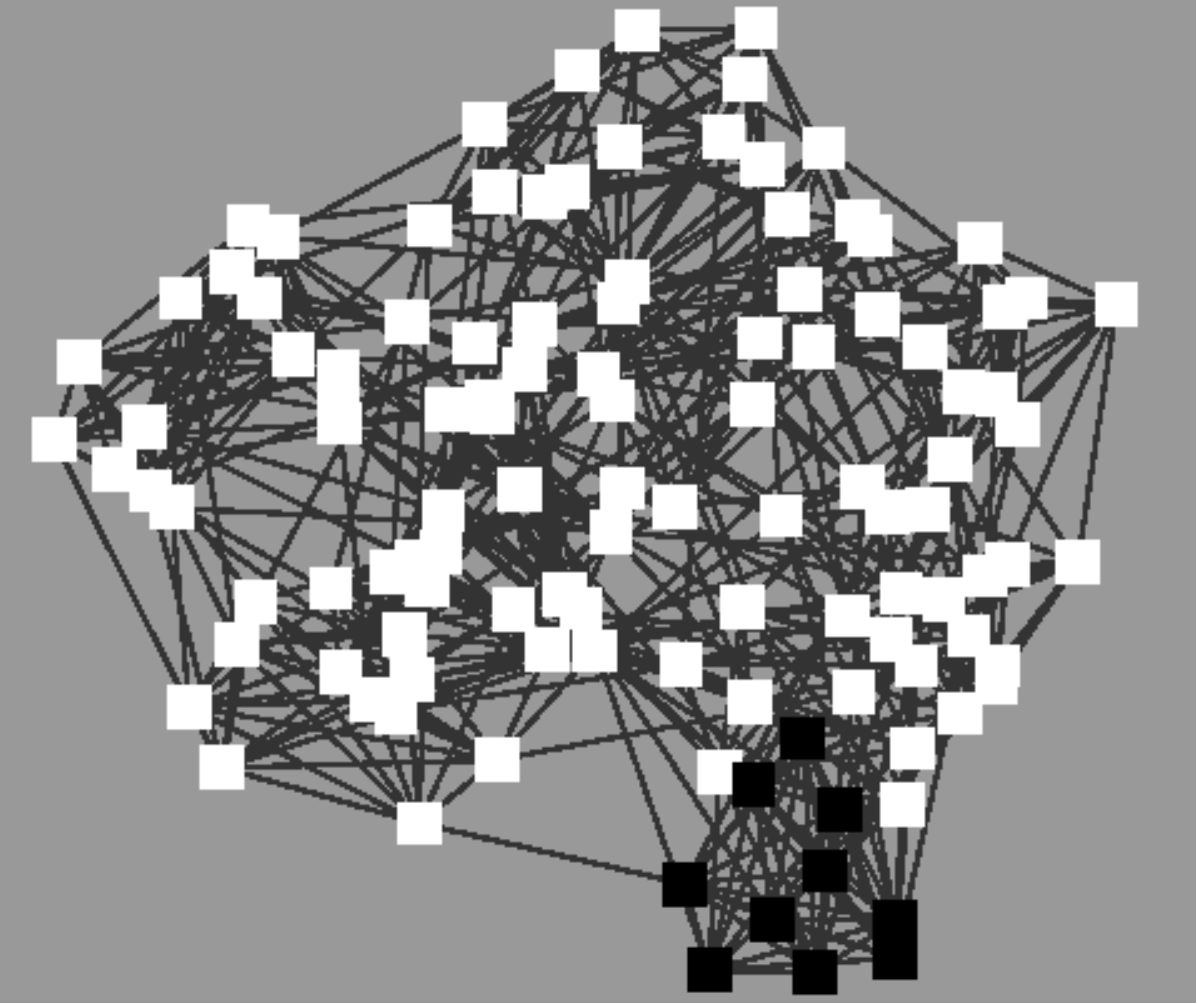}}
\hspace{0mm}
\subfigure[]
{\includegraphics[width=4cm]{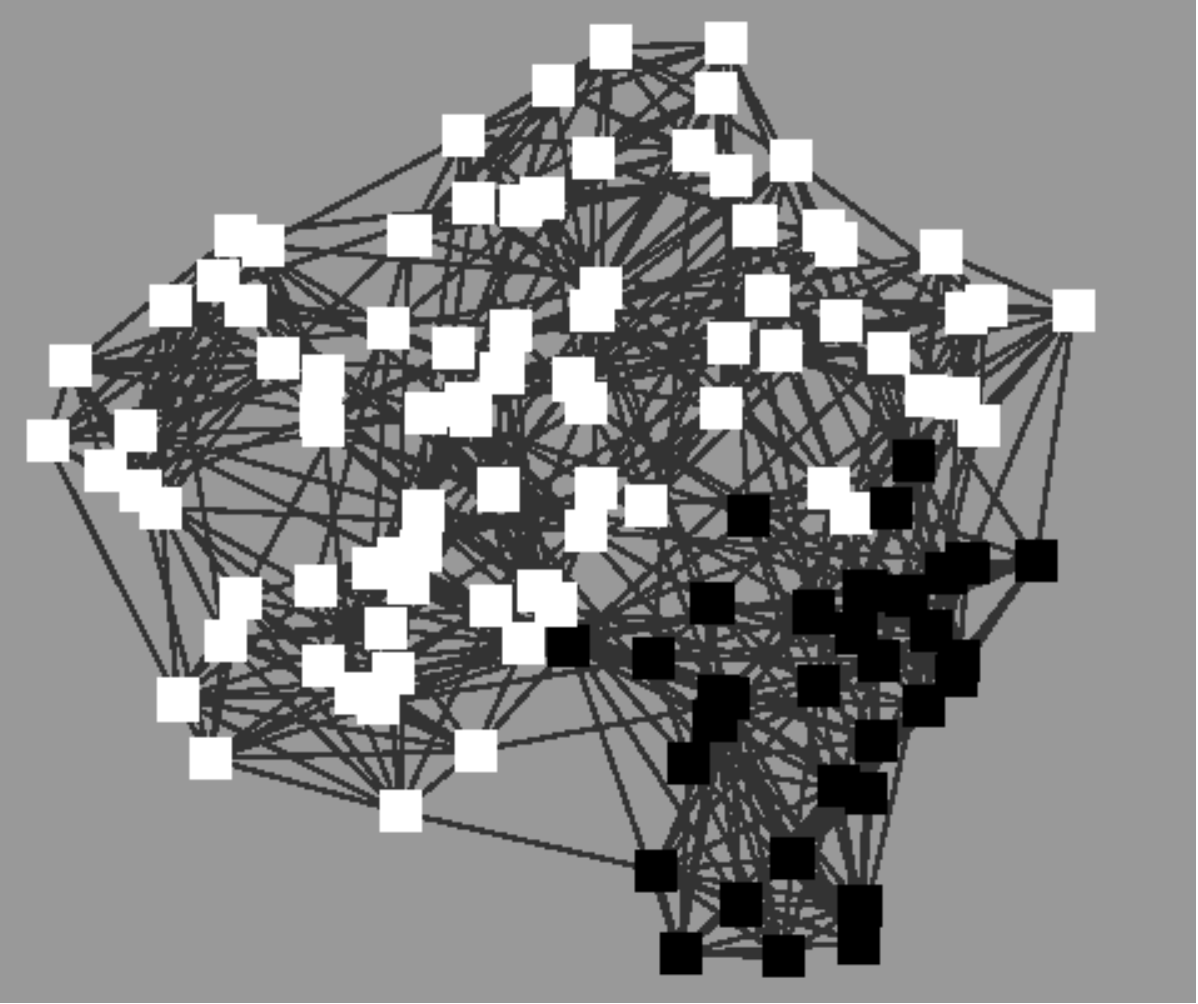}}
\hspace{0mm}
\subfigure[]
{\includegraphics[width=4cm]{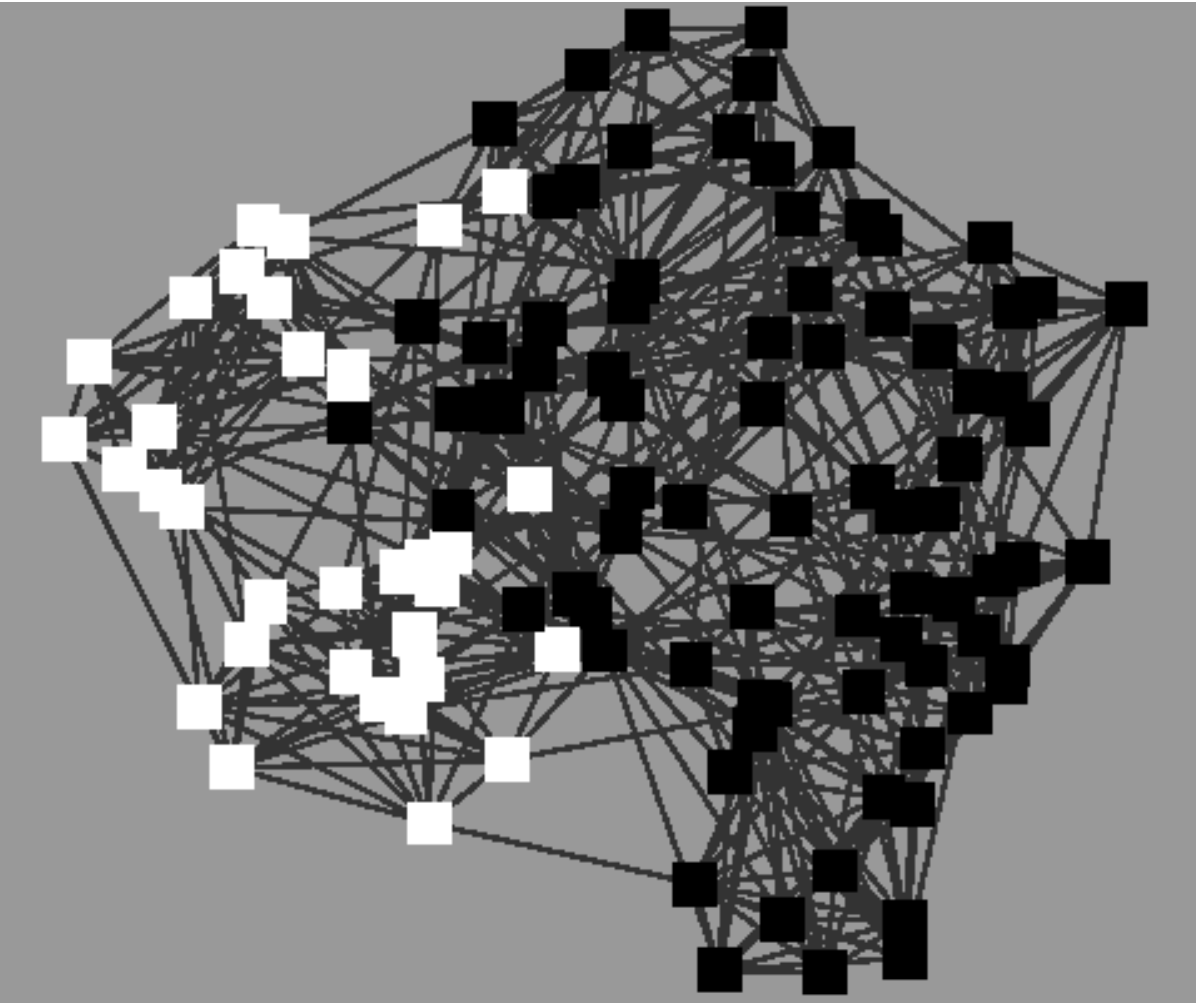}}
\hspace{0mm}
\subfigure[]
{\includegraphics[width=4cm]{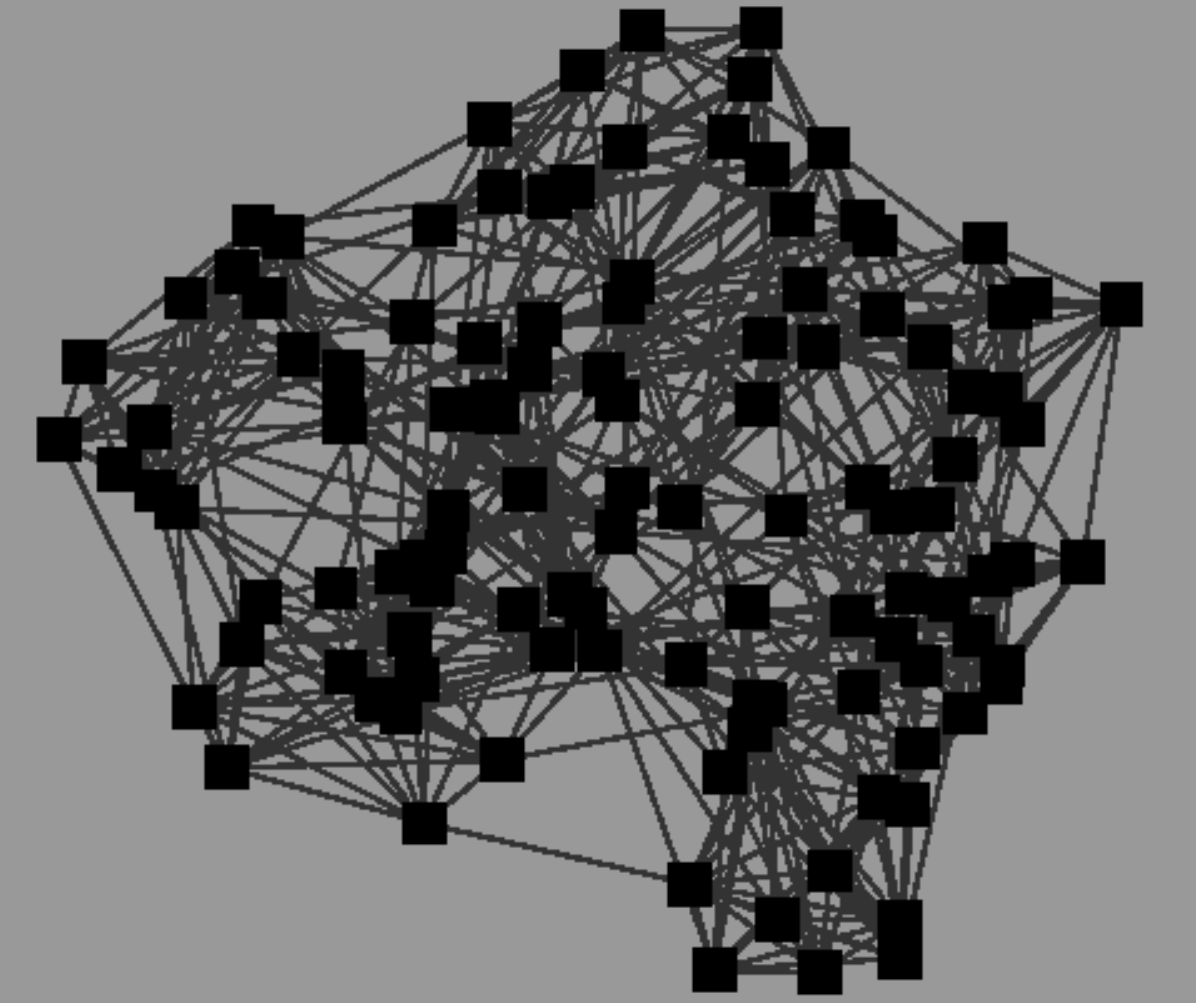}}
\caption{\label{fig:foot} Washington (black node in $1$) in NCAA football college network. Black nodes indicate the community of \emph{Washington} for different time steps.}
\end{figure} 

\begin{figure}[t]
\begin{center}
\includegraphics[width=.6\columnwidth]{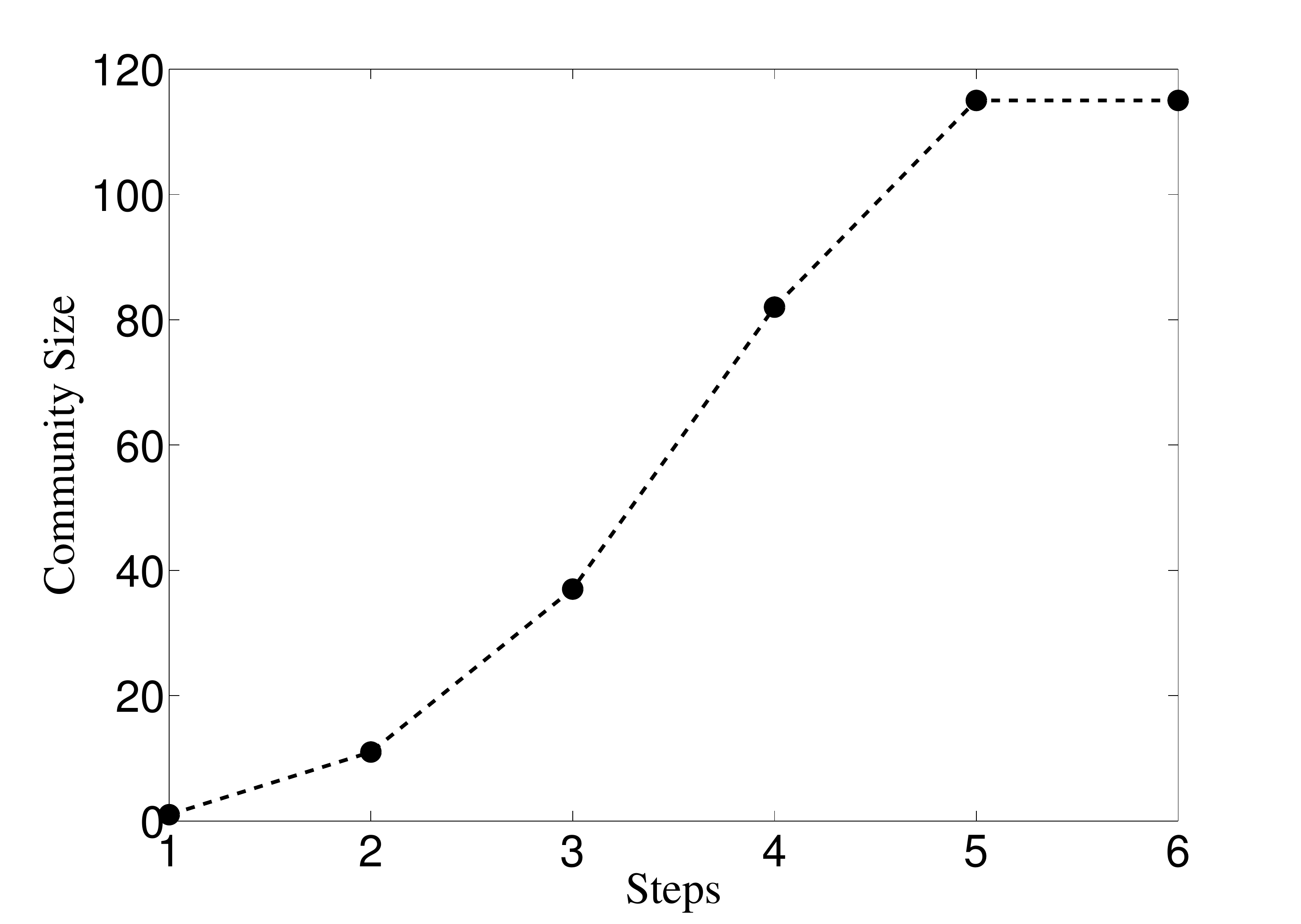} 
\end{center}
\caption{Temporal evolution (\emph{x-axis}) of community size ($y-axis$) for Washington in the \emph{NCCA College Football} network. }
\label{fig:foot_temp}
\end{figure}

\section{Improvement of the method}
\label{improve}
In this section we want to explore the performances of our method taking into account two different heuristics in order to free the method from the precise tuning of the parameters $m$ and $\alpha$. The main feature of these two heuristics called respectively \emph{IDA + LTE} and \emph{Double pruning} is to reveal  the hierarchical community structure of complex networks from a local viewpoint. 
\subsection{IDA + LTE}
Recently Huang et al.~\cite{Huang11} proposed a local algorithm for detecting communities in networks. Here we report a general description of this method, known as Local Tightness Expansion Expansion (LTE), while it is well described in the Appendix. This method can reveal the community from a starting vertex via local optimization of the tightness measure which is proposed in Ref~\cite{Radicchi04}, but the expansion is given by a tunable parameter which is not ecological in social networks where people elaborate the information trough mental scheme or cognitive heuristics. The LTE algorithm can be summarized in the following: 
\begin{enumerate}
\item Pick a vertex $s \in V $ as the starting vertex. \\
Let $C =  \left\{S \right\} $ and $N = \Gamma(S) -\left\{S\right\}$
\item Select the vertex \textit{{\large a}} $\in N$ that possess the largest
similarity with vertices in $C$.
\item If $ \tau{_C}^{\beta}(A) >0$, \\
set $C = C \cup  \left\{a \right\} $ and $N = N \cup \Gamma (a) - C $
\item Repeat 2 and 3 until $N= \varnothing$.
\end{enumerate}

The LTE  algorithm,  is  very efficient for detecting communities in networks from an individual point of view. When $N= \varnothing$ we arrive at a static convergence. Here we want to introduce a new method merging together our algorithm, that we call Information Dynamics Algorithm (IDA), and LTE for analysing how the information dynamics can improve the performance of LTE and vice versa. We describe the new algorithm in the following (its representation is illustrated in Fig.~\ref{fig:nuovoalgortimschema}):
\begin{enumerate}
\item Pick a vertex $s \in V $.
\item Run LTE and discover $C_{LTE}$ for the vertex $s$.
\item Run IDA and take information of other nodes by community nodes
\item Run LTE and decide to accept new nodes. 
\item Repeat 3 and 4 until a fixed memory or when the vertex $s$ knows the whole network. 
\end{enumerate}

For testing purposes we use two real networks  analyzing and discussing our model peculiarities. The two case studies, of growing or different complexity, are  the  Zachary \emph{karate club} network~\cite{Zachary} as reported in the Fig.~\ref{fig:karate_completa} and Fig.~\ref{fig:karate_temp} and the NCAA college football network~\cite{dolphin} as shown in the Fig.~\ref{fig:foot} and Fig.~\ref{fig:foot_temp}. For all the simulations we used always the same parameters. For the IDA we assumed $\alpha = 1.4$ and $m = 0.2$. For the LTE we used $\beta = 0.3$ for the first step and $\beta = 2$ during LTE+IDA dynamics.
As shown in these figures, our algorithms is able to discover the multi-resolution community structure of each node detecting the different local inner circles~\cite{dunbar, dunbar2}.

\subsection{Double pruning}
\begin{figure}[t!]
\centering
\subfigure[]
{\includegraphics[width=4cm]{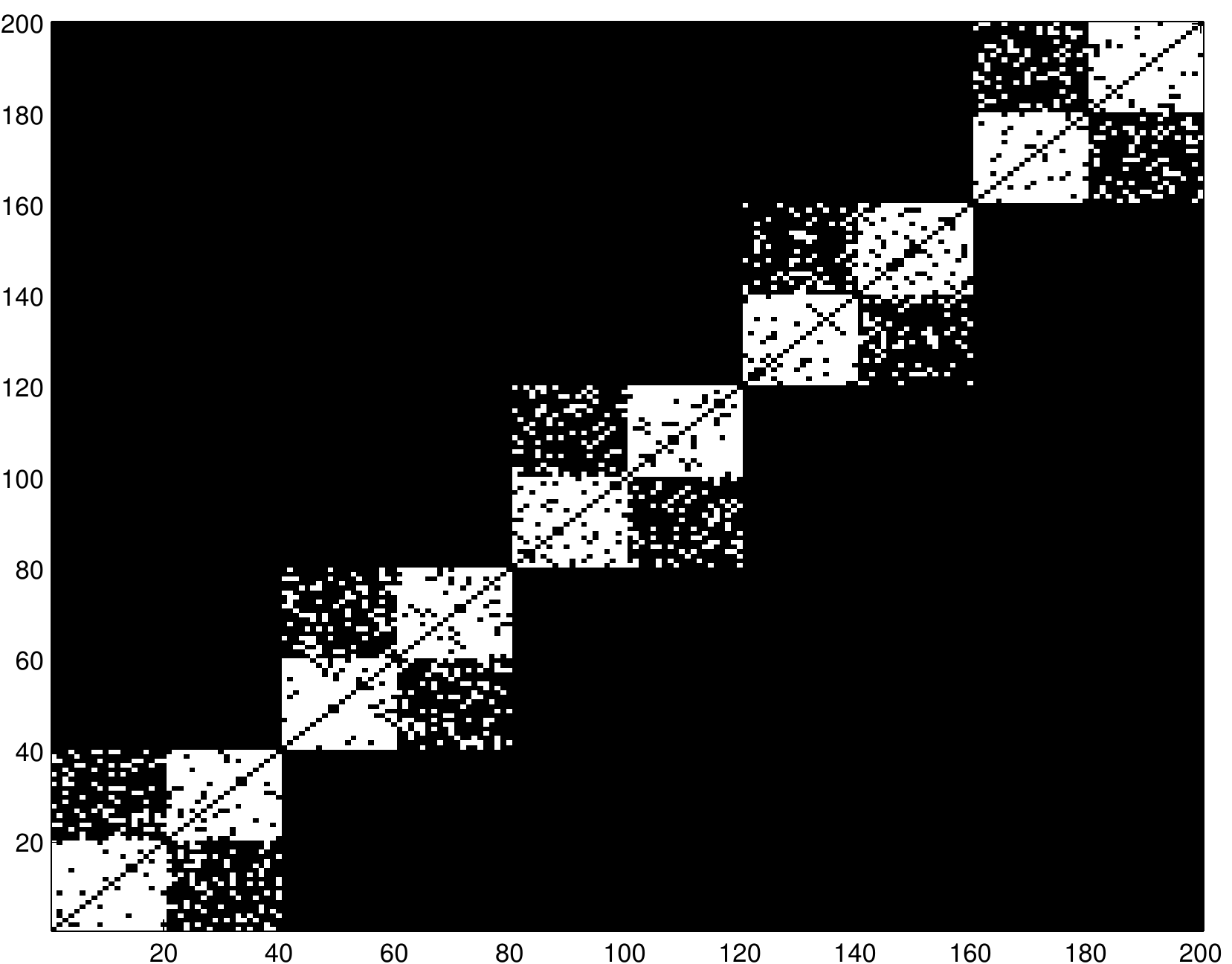}}
\hspace{0mm}
\subfigure[b=2]
{\includegraphics[width=4cm]{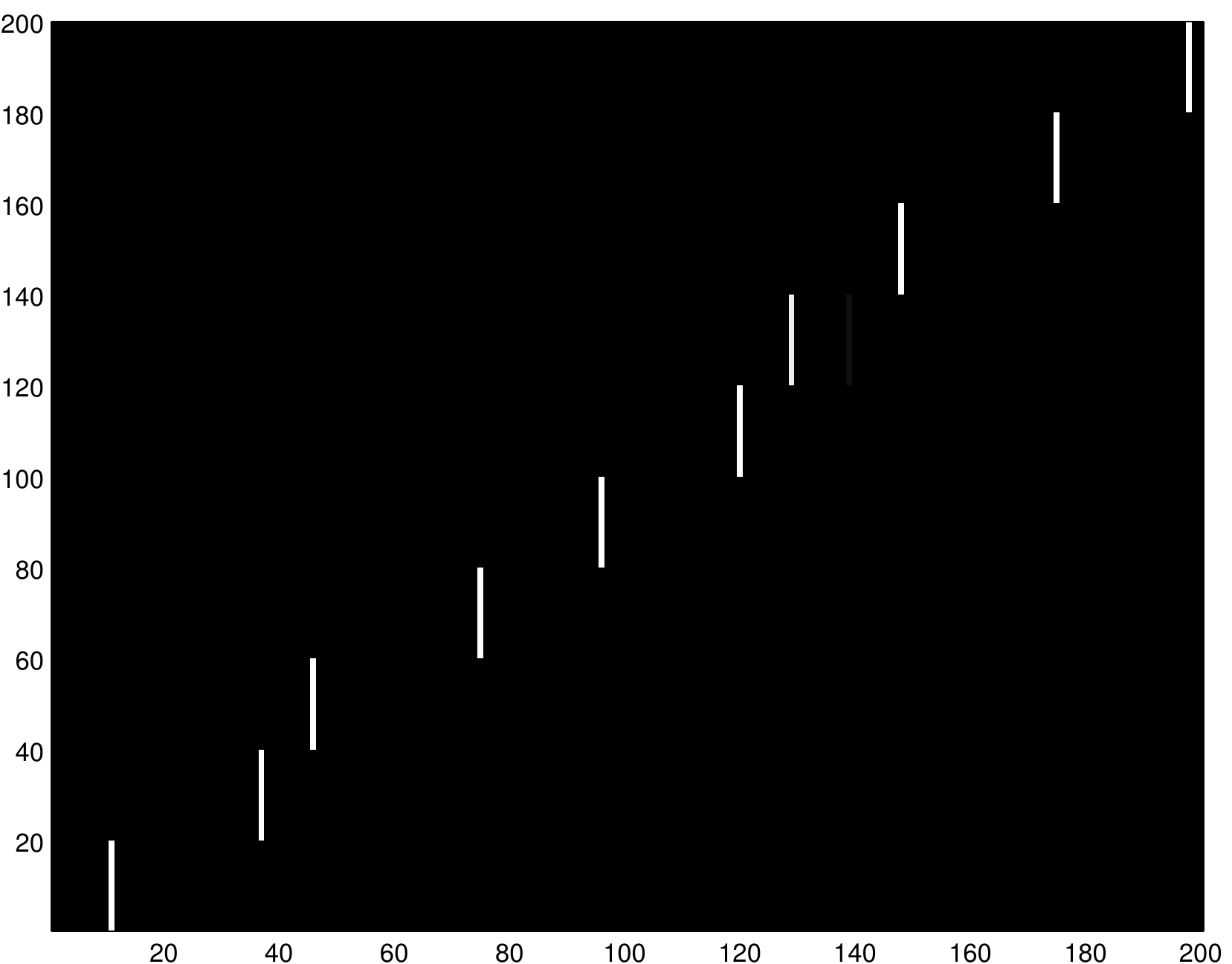}}
\hspace{0mm}
\subfigure[b=4]
{\includegraphics[width =4cm]{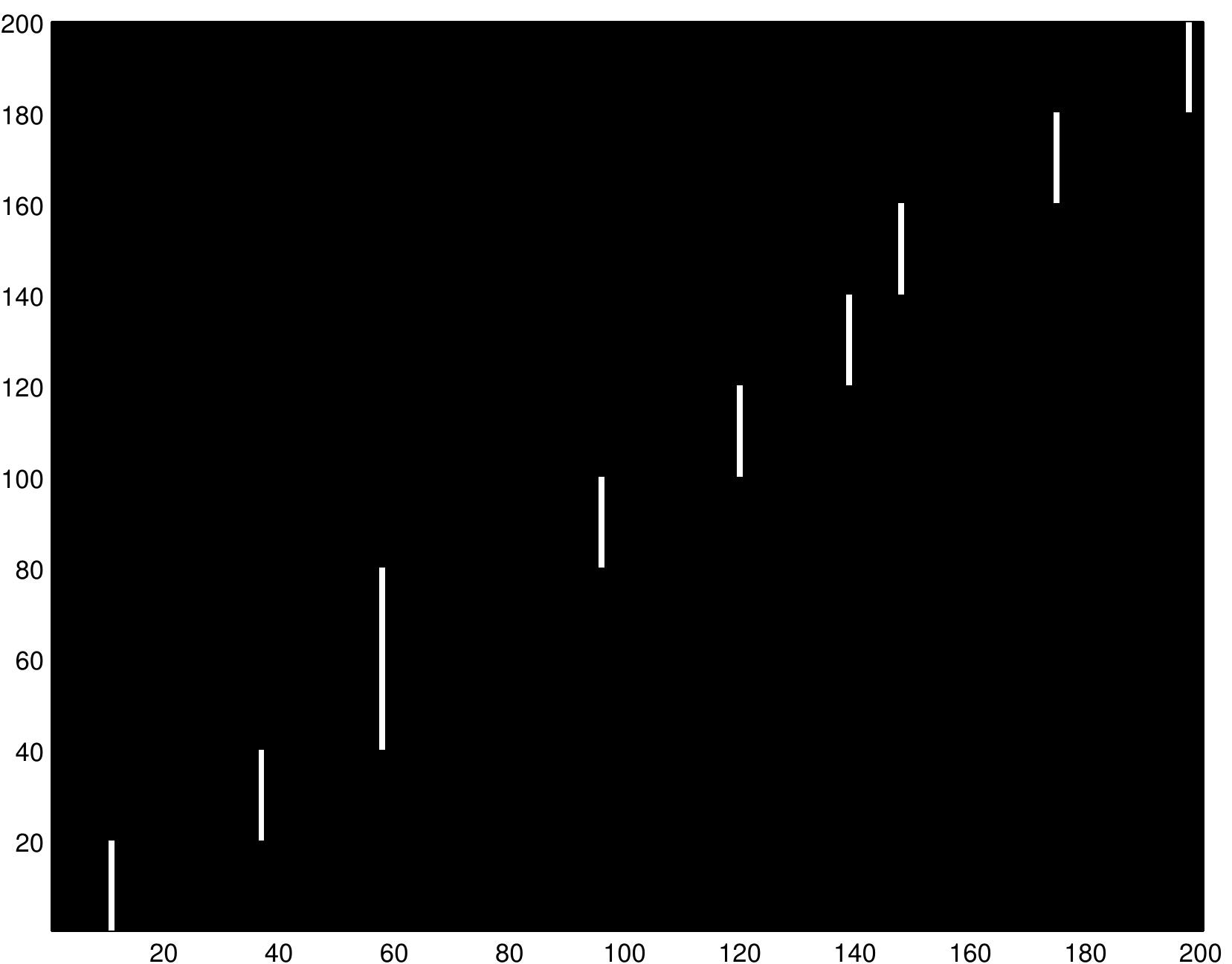}}
\hspace{0mm}
\subfigure[b=6]
{\includegraphics[width =4cm]{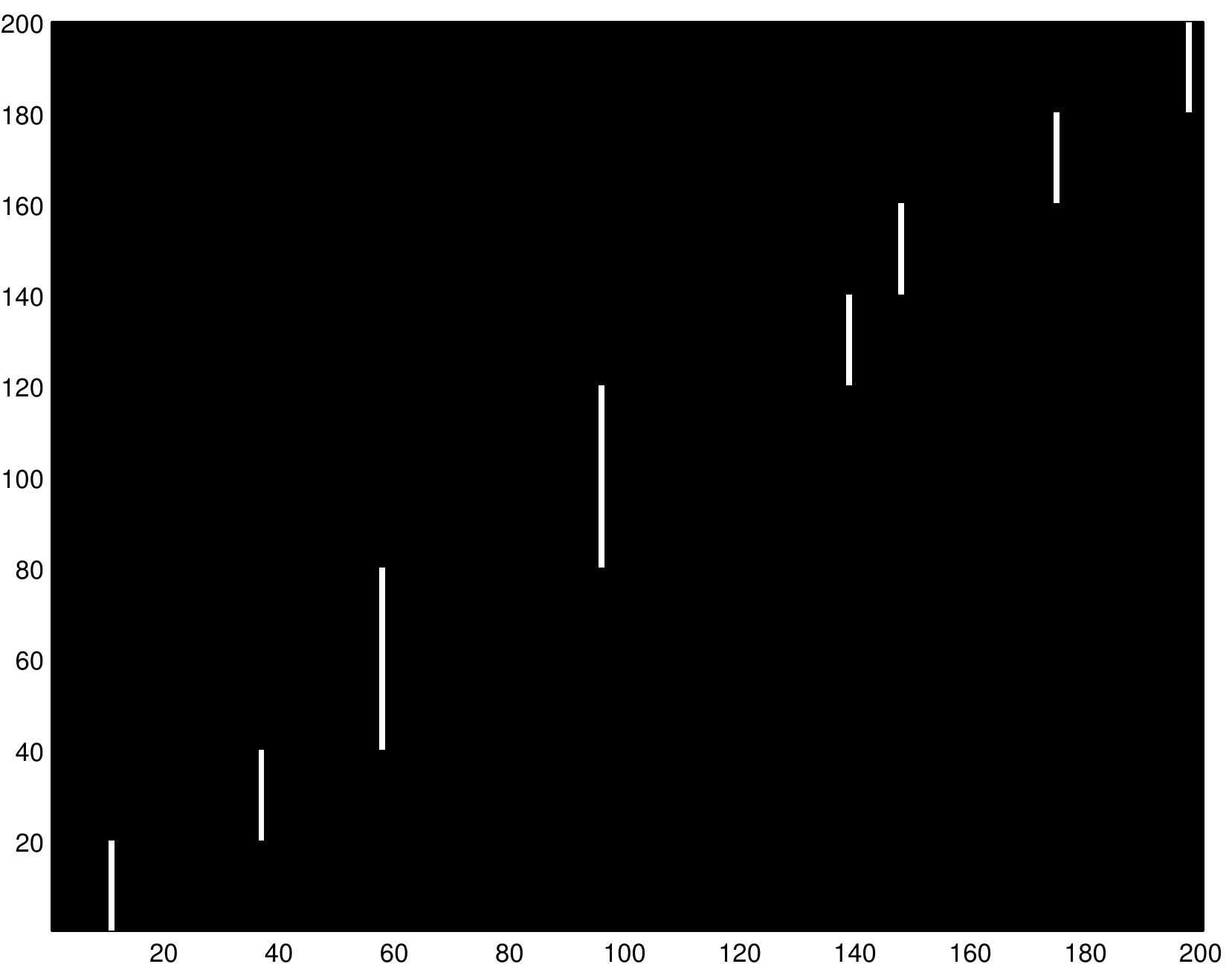}}
\hspace{0mm}
\subfigure[b=8]
{\includegraphics[width =4cm]{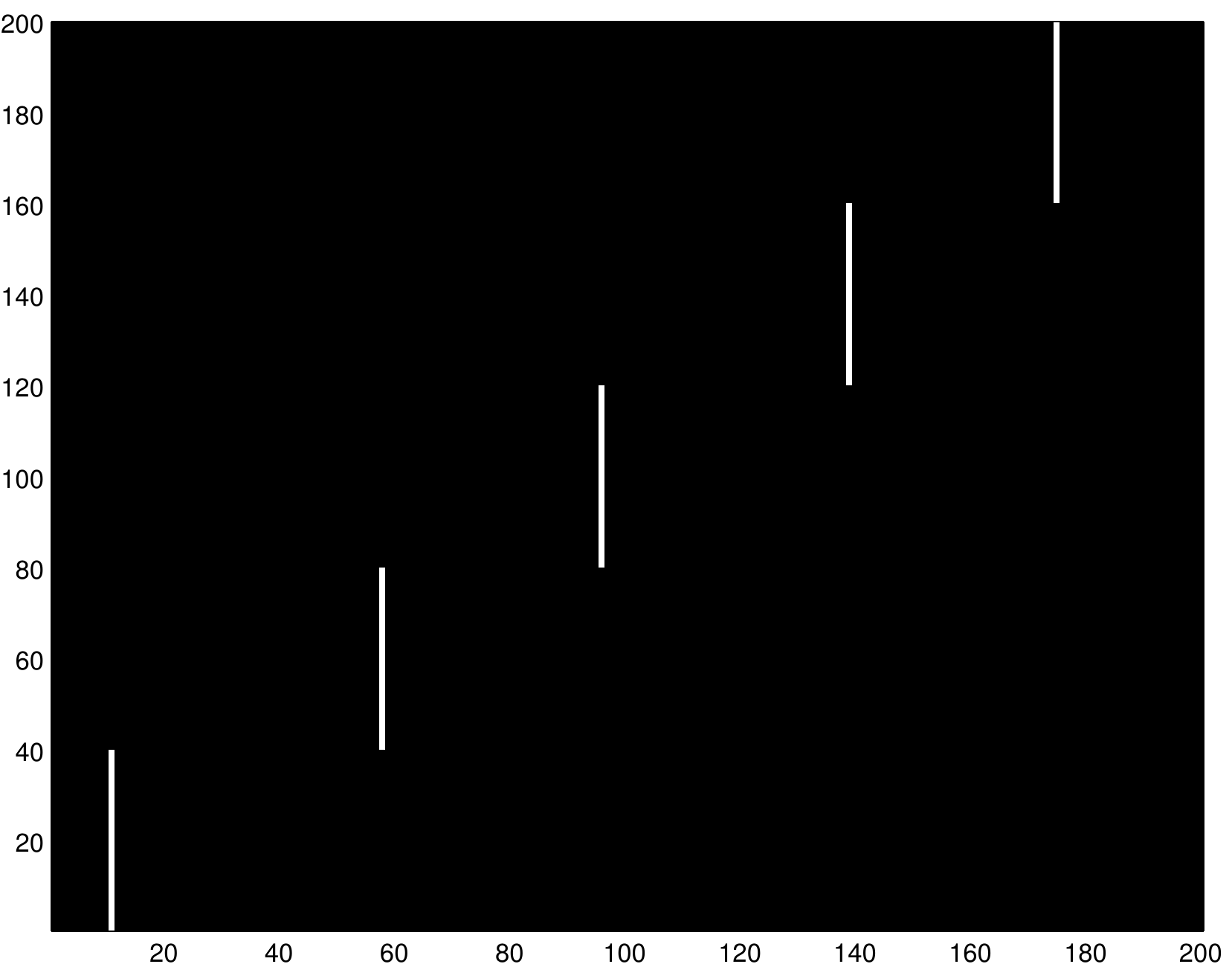}}
\caption{\label{fig:fig1} (a) Adjacency matrix of a network composed by $200$ nodes and $3$ levels where $p_1=0.9$, $p_2=0.2$ and $p_3=0$. Trough the double pruning heuristic it is possible to detect the different community levels reaching the discover of the four principal communities for $b=8$.}
\end{figure}

\begin{algorithm}[t!]
\algsetup{linenosize=\tiny}
\scriptsize
\begin{algorithmic}[1]
\STATE $T$, $T_{max}$, $t_b=0$, $t_1=0$, $c=0$, $a=0$
\STATE $m=0.2$, $\alpha=1.4$ 
 \WHILE{$t < T$} 
 \STATE{$t=t+1$; $b = b+1$;} 
  \IF{$b<2$} \STATE{$t_{max} = b*50$} \ELSE \STATE{$t_{max}=b*10$} \ENDIF
  \FOR{$t _1 = 1:t_{tmax}$} 
  \STATE{$a = a + 1;$} 
  \IF{$a=b$} 
  \STATE{we evaluate the state vector of each node, then $S^1$ is vector state of node 1, $S^2$ of node 2 and so on...} 
  \FOR{$i = 1:N$}
  \STATE{$ dx^i = (S^i_{max} - S^i_{min})/3$}
  \FOR{$j = 1:N$}
  \IF{$S_{i,j}<S^i_{min}+2*dx^i$}
  \STATE{$S_{i,j}=0$}
  \ENDIF
  \ENDFOR
  \ENDFOR
  \STATE{we normalize $S$}
  \ELSE 
  \STATE{$S_1 = mS_0 + (1-m)M'S_0$, $S_2 = S_1^{\alpha}$}
  \STATE{we normalize $S$}
  \STATE{a=0}
   \ENDIF
  \ENDFOR
  \STATE{$S = I(N)$}
\ENDWHILE
\end{algorithmic}
\caption{pseudocode for double pruning heuristic }
\label{alg:double}
\end{algorithm}
\begin{figure}[t!]
\centering
\subfigure[]
{\includegraphics[width=1\columnwidth]{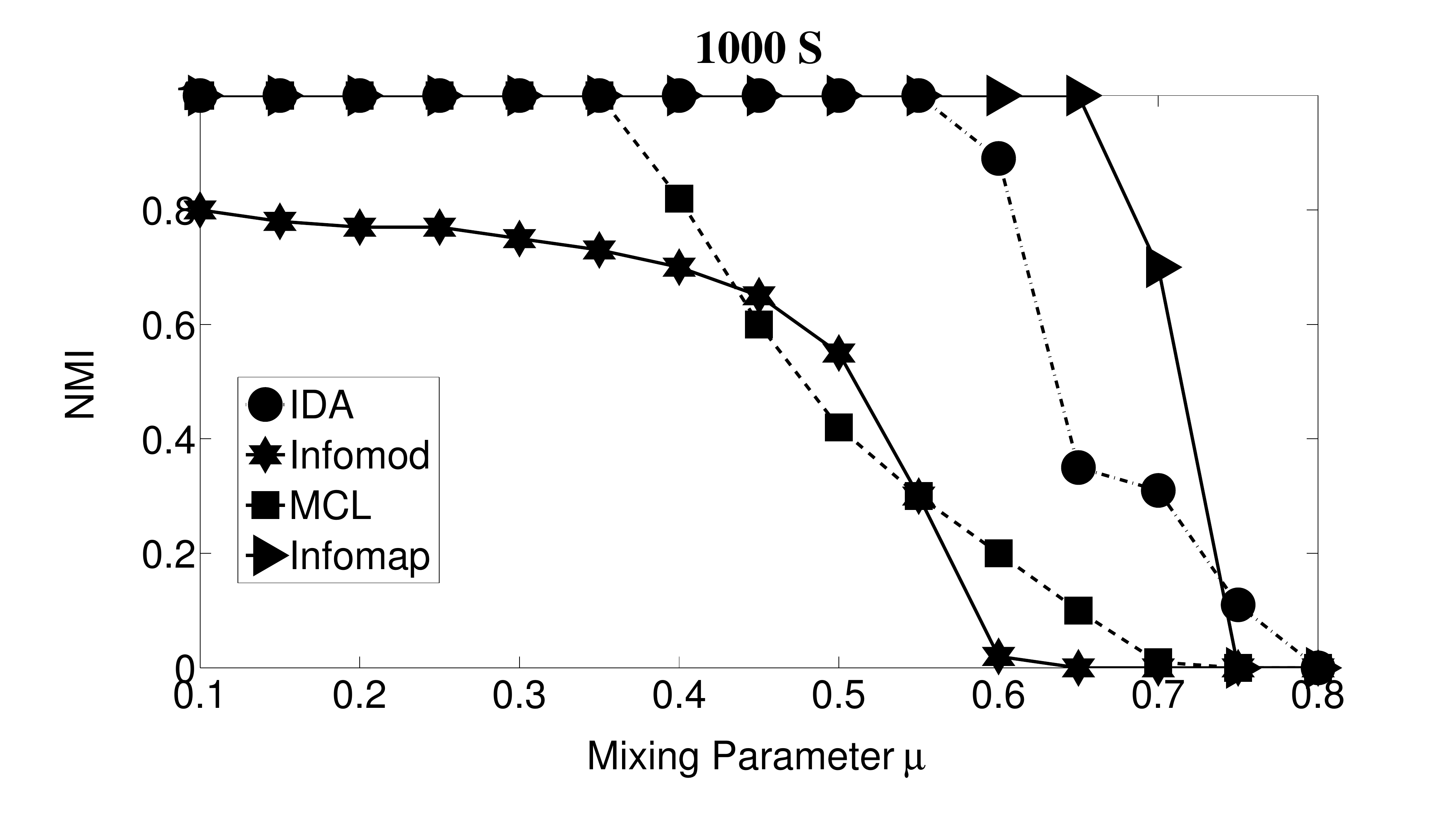}}
\hspace{0mm}
\subfigure[]
{\includegraphics[width=1\columnwidth]{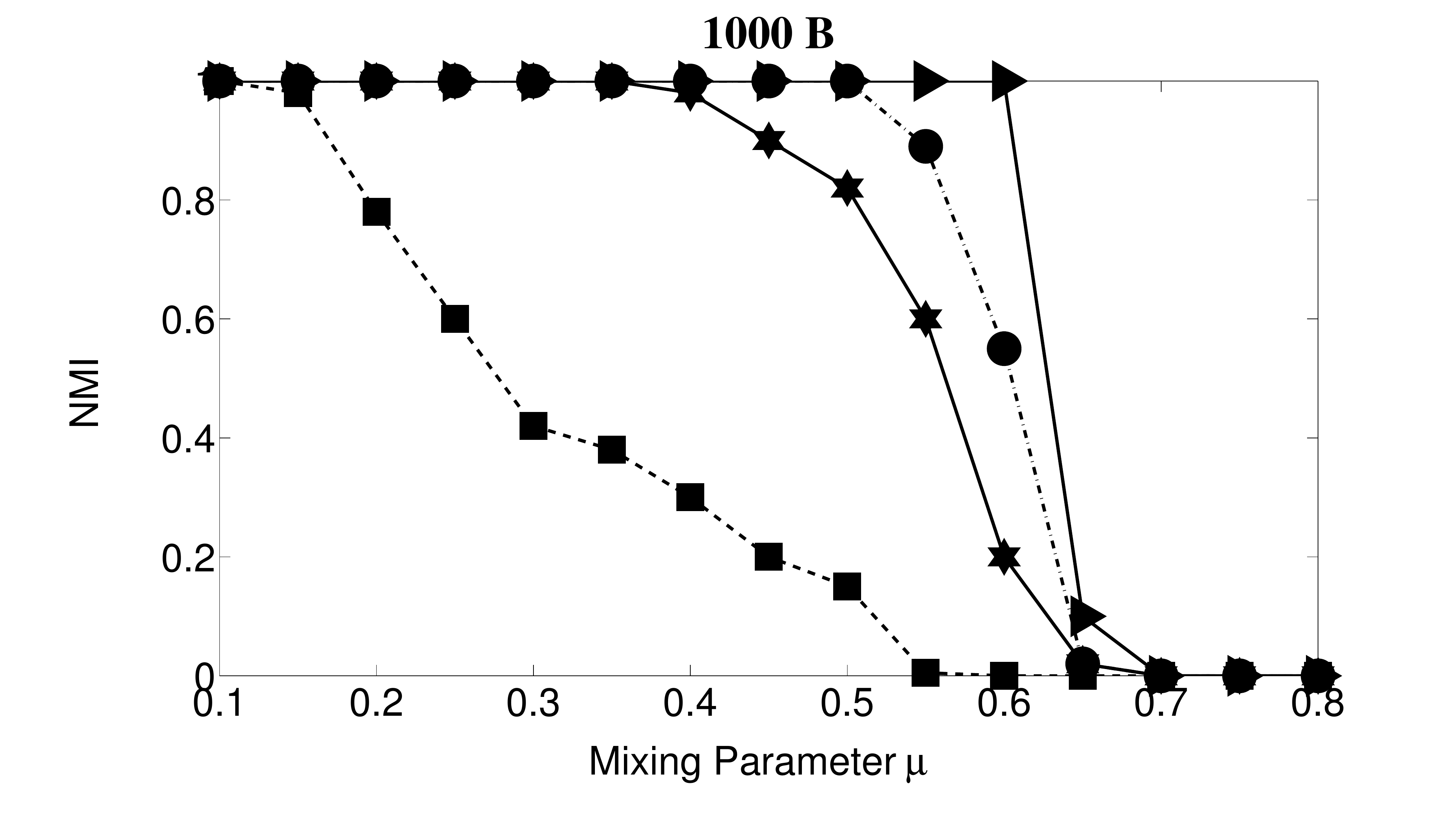}}
\caption{\label{fig:nmi} (a)-(b) Comparison between our method (IDA) and other algorithms (Infomod, MCL and Infomap) on the LFR Benchmark graphs with different size of communities respectively Small~(a) and Big~(b) increasing the mixing parameter $\mu$.}
\end{figure} 

Our model has two free parameters $m$ and $\alpha$ and in previous works we have shown that it is very difficult to find the specific values of these parameters needed to have a good representation of community distribution for different networks~\cite{Massaro2012, Bagnoli2012, Guazzini2012}. However, there is a simple procedure that can be used:  at some moments for each node we evaluate the histogram of the state vector with $3$ bins and we set to zero the values that are lower then the second bin values, as reported in the Algorithm~\ref{alg:double}. 

In this way we are able to generate different view of the clustering levels on hierarchical community structure networks as shown in Fig.~\ref{fig:fig1}. We use the Lancichinetti-Fortunato-Radicchi (LFR) benchmark graph~\cite{lancichinetti2009community, LFR} to evaluate the accuracy of this method. We adopt the normalized mutual information~(NMI) to evaluate the quality of detected communities which is currently widely used in measuring the performance of graph clustering algorithms~\cite{lancichinetti2009community}.
The accuracy of our method is compared with three others well-known community detection algorithms. The observed results on our hierarchical networks allow us to state that for $b=4$ we are able to detect the principal communities. For this reason for comparing our method with others we set $b=4$. 

In order to evaluate and partially validate our approach, we have applied our algorithm comparing its performance with $3$ community detection algorithms namely \textit{Infomap}~\cite{infomap}, \textit{Infomod}~\cite{infomod} and \textit{MCL}~\cite{MCL} as reported in Fig.\ref{fig:nmi}. The input parameters of the benchmark graphs used here are: number of nodes $N=1000$, average degree is $\hat{k}=20$,  maximum degree $\hat{k}=50$ for all the networks. Moreover we changed the range of community size generating two kinds of networks of $1000$ nodes: $1000S$ ($S$ stay for small)  means that communities have between 10 and 50 nodes  and $1000B$ ($B$ stay for big) means that communities have between $20$ and $100$ nodes. Results of the performance's comparison between our algorithm and the others are reported in Fig.~\ref{fig:nmi}. Our algorithm with the double pruning heuristic is very competitive with the other algorithms except for the Infomap method which is nowadays the best algorithm for detecting communities in static network, even if it can not be easily applied in dynamic environments.
\section{Conclusions}  
In this paper we presented two extensions of a community detection algorithm which is based on pure local information propagation inspired by the so-called Markov Clustering Algorithm (MCL) by S. Van Dongen~\cite{MCL}. The proposed method is able to identify  both overlapping and non-overlapping communities, and we have shown that it is capable to identify communities also in dynamic networks~\cite{Massaro2012, Bagnoli2012, Guazzini2012}. Our algorithm proved to be a multi-level solution that can be used to capture the hierarchical community structure from a local viewpoint at any resolution. Experimental results on the real-world and synthetic datasets show that our algorithm achieves good performance. Concluding, we showed how the adaptation of two heuristics, namely \emph{IDA+LTE} and \emph{Double Pruning} are  effective in freeing the method  of the precise tuning of parameters. 

\bibliography{biblio_review}

\begin{thebibliography}{10}

\bibitem{Massaro2012}
E.~Massaro, F.~Bagnoli, A.~Guazzini, and P.~Li\'o, ``Information dynamics
  algorithm for detecting communities in networks,'' {\em Comm. Nonlin. Sci
  Numer. Simul.}, vol.~17, no.~11, pp.~4294 -- 4303, 2012.

\bibitem{Erdos}
P.~Erdős and A.~Rényi, ``On the evolution of random graphs,'' in {\em
  Publication of the Mathematical Institute of the Hungarian Academy of
  Sciences}, pp.~17--61, 1960.

\bibitem{Barabasi99}
A.~L. Barabasi and R.~Albert, ``Emergence of scaling in random networks,'' {\em
  Science}, vol.~286, no.~5439, pp.~509--512, 1999.

\bibitem{Newman2002}
M.~E.~J. Newman, ``Assortative mixing in networks,'' {\em Phys. Rev. Lett.},
  vol.~89, no.~20, p.~208701, 2002.

\bibitem{albert1999}
R.~Albert, H.~Jeong, and A.~L. Barabasi, ``{The diameter of the world wide
  web},'' {\em Nature}, vol.~401, pp.~130--131, 1999.

\bibitem{Watts1998}
D.~J. Watts and S.~H. Strogatz, ``Collective dynamics of {'}small-world{'}
  networks,'' {\em Nature}, vol.~393, pp.~440--442, June 1998.

\bibitem{ravasz2003}
E.~Ravasz and A.~L. Barab\'asi, ``Hierarchical organization in complex
  networks,'' {\em Phys. Rev. E}, vol.~67, no.~2, p.~026112, 2003.

\bibitem{Girvan02}
M.~Girvan and M.~E.~J. Newman, ``Community structure in social and biological
  networks.,'' {\em PNAS}, vol.~USA 99, p.~7821–7826, 2002.

\bibitem{Fortunato}
S.~Fortunato, ``Community detection in graphs,'' {\em Physics Reports},
  vol.~486, no.~3–5, pp.~75 -- 174, 2010.

\bibitem{Newman:2004fk}
M.~E.~J. Newman, ``Detecting community structure in neworks,'' {\em The
  European Physical Journal B}, vol.~38, no.~2, pp.~321--333, 2004.

\bibitem{Palla}
G.~Palla, I.~Derényi, I.~Farkas, and T.~Vicsek, ``Uncovering the overlapping
  community structure of complex networks in nature and society,'' {\em
  Nature}, pp.~814--818, June.

\bibitem{Bagrow2008}
J.~P. Bragrow, ``Evaluating local community methods in networks,'' {\em Journal
  of Statistical Mechanics: Theory and Application}, p.~P05001, 2008.

\bibitem{Bagrow20081}
J.~P. Bagrow and E.~M. Bollt, ``Local method for detecting communities,'' {\em
  Phys. Rev. E}, vol.~72, p.~046108, Oct 2005.

\bibitem{Clauset2005}
A.~Clauset, ``Finding local community structure in networks,'' {\em Phys. Rev.
  E}, vol.~72, p.~026132, Aug 2005.

\bibitem{Luo2008}
F.~Luo, J.~Wang, and E.~Promislow, ``Exploring local community structures in
  large networks,'' in {\em Web Intelligence, 2006. WI 2006. IEEE/WIC/ACM
  International Conference on}, pp.~233 --239, dec. 2006.

\bibitem{Lanc09}
A.~Lancichinetti, S.~Fortunato, and J.~Kertész, ``Detecting the overlapping
  and hierarchical community structure in complex networks,'' {\em New Journal
  of Physics}, vol.~11, no.~3, p.~033015, 2009.

\bibitem{Huang11}
J.~Huang, H.~Sun, Y.~Liu, Q.~Song, and T.~Weninger, ``Towards online
  multiresolution community detection in large-scale networks,'' {\em PLoS
  ONE}, vol.~6, no.~8, p.~e23829, 2011.

\bibitem{LFR}
A.~Lancichinetti, S.~Fortunato, and F.~Radicchi, ``Benchmark graphs for testing
  community detection algorithms,'' {\em Phys. Rev. E}, vol.~78, p.~046110,
  Oct. 2008.

\bibitem{MCL}
S.~V. Dongen, {\em {Graph Clustering by Flow Simulation}}.
\newblock PhD thesis, University of Utrecht, 2000.

\bibitem{Radicchi04}
F.~Radicchi, C.~Castellano, F.~Cecconi, V.~Loreto, and D.~Parisi, ``Defining
  and identifying communities in networks,'' {\em PNAS}, vol.~101, p.~2658,
  2004.

\bibitem{Zachary}
W.~W. Zachary, ``An information flow model for conflict and fission in small
  groups.,'' {\em Journal of Anthropological Research}, no.~33, p.~452–473,
  1977.

\bibitem{dolphin}
D.~Lusseau, K.~Schneider, O.~J. Boisseau, P.~Haase, E.~Slooten, and S.~M.
  Dawson {\em Behavioral Ecology and Sociobiology}, no.~54, p.~396–405, 2003.

\bibitem{dunbar}
R.~Dunbar, ``Neocortex size as a constraint on group size in primates,''

\bibitem{dunbar2}
R.~Dunbar, ``Coevolution of neocortical size, group size and language in
  humans,''

\bibitem{Bagnoli2012}
F.~Bagnoli, E.~Massaro, and A.~Guazzini, ``Community-detection cellular
  automata with local and long-range connectivity.,'' {\em LNCS, Springer,
  Berlin 2012}, vol.~7495, pp.~204--213, 2012.

\bibitem{Guazzini2012}
D.~Borkmann, A.~Guazzini, E.~Massaro, and S.~Rudolph, ``A cognitive-inspired
  model for self-organizing networks,'' in {\em IEEE Sixth International
  Conference on Self-Adaptive and Self-Organizing Systems Workshops (SASOW)},
  pp.~229--234, 2012.

\bibitem{lancichinetti2009community}
A.~Lancichinetti and S.~Fortunato, ``Community detection algorithms: a
  comparative analysis,'' 2009.

\bibitem{infomap}
M.~Rosvall and C.~T. Bergstrom, ``Maps of random walks on complex networks
  reveal community structure,'' {\em PNAS}, vol.~105, no.~4, pp.~1118--1123,
  2008.

\bibitem{infomod}
M.~Rosvall and C.~Bergstrom, ``{An information-theoretic framework for
  resolving community structure in complex networks},'' {\em Proceedings of the
  National Academy of Sciences}, vol.~104, no.~18, p.~7327, 2007.

\end{thebibliography}
\bibliographystyle{ieeetr}
\clearpage

\appendix
\pagestyle{empty}
\section{Appendix I}
Usually, a network can be represented by a graph $G=(V,E)$, where $V$ is the set of vertices and $E$ is the set of edges.
{\bf Definition 1} \textit{(Neighborhood) Let~~ $G=(V,E,w)$ be a weighted undirected network and $w(e)$ be the weight of the edge $e$. For a vertex $u\in V$, the structure neighbourhood of vertex $u$ is the set $\Gamma(u)$ containing $u$ and its adjacent vertices which are incident with a common edge with $u$ : $\Gamma(u) = \left\{ v \in V  | \left\{u,v\right \} \in E \right \} \cup \left\{u \right \}$.}

{\bf Definition 2} \textit{(Structural Similarity) Given a weighted undirected network $G=(V,E,w)$, the structure similarity $s(u,v)$ between two adjacent vertices $u$ and $v$ is:}
\begin{equation}  
 s(u,v) = \frac{\displaystyle\sum_{x \in \Gamma(u) \cap \Gamma(v)}  w(u,x) \cdot w(v,x)}    {\displaystyle\sqrt{\sum_{x \in \Gamma(u)} w^2(u,x)}\cdot  {\displaystyle\sqrt{\sum_{x \in \Gamma(v)} w^2(v,x)}}  }. 
\end{equation}

When we consider an unweighted graph, the weight $w(u,v)$ of any edge $\left\{u,v\right \} \in E$ can be set to 1 and the equation above can be transformed to:
\begin{equation}  
 s(u,v) = \frac{|\Gamma(u) \cap \Gamma(v)|}{\sqrt{|\Gamma(u)| \cdot |\Gamma(v)|}}. 
\end{equation}

It corresponds to the so-called edge-clustering coefficient introduced by Radicchi et al.~\cite{Radicchi04}.

{\bf Definition 3} \textit{(Tightness) By employing the structural similarity, we introduce tightness, a new quality function of a local community $C$, which is given as follows:
\begin{equation}  
T(C) = \frac{S_{in}^C}{S_{in}^C + S_{out}^C} 
\end{equation}
where $S_{in}^C = {\displaystyle\sum_{u \in C, v \in C, \left\{u,v\right \} \in E }  s(u,v)}$
 is the internal similarity of the community $C$ which is equal to two times of the sum of similarities between any two  adjacent vertices both inside the community C; \\
$S_{out}^C = {\displaystyle\sum_{u \in C, v \in N, \left\{u,v\right \} \in E }  s(u,v)}$ is the external similarity of the community C which is equal to the sum of
similarities between vertices inside the community C and vertices out of it.}

The tightness measure is extended from the weak community
definition proposed in Ref.~\cite{Radicchi04}. Similar to other community definitions, the tightness value of a community
$C$, denoted by $T(C)$, will increase when sub-graph $C$ has high internal similarity and low external similarity. The whole network
without outward edges will achieve the maximal value 1, but the problem here is to find the local optimization of the measurement
for each community. Suppose a community $C$ is detected from a certain vertex $s$. We explore the adjacent vertices in the neighbourhood set $N$ of $S$ as
shown in Fig. 2. So the variant tightness of the community $C \cup \left\{A\right \}$ becomes
\begin{equation} 
\begin{split} 
T(C \cup \left\{A\right \}) &= \frac{S_{in}^C + 2S_{in}^a}{(S_{in}^C + S_{in}^a)+(S_{out}^c - S_{in}^a + S_{out}^a )} =\\
&=\frac{S_{in}^C + 2S_{in}^a}{(S_{in}^C + S_{in}^a+S_{out}^c + S_{out}^a)},
\end{split}
\end{equation}
where $S_{in}^a = {\displaystyle\sum_{ \left\{v,a\right \} \in E \wedge v \in C}  s(v,a)} $; $S_{out}^a = {\displaystyle\sum_{ \left\{a,u\right \} \in E \wedge u \not\in C}  s(a,u)} $. Then the tightness increment of a vertex \textit{{\large a}} joining in $C$ is:
\begin{equation} 
\begin{split} 
\Delta T_C(A) &= T(C \cup \left\{A\right \}) =\\
&=\frac{S_{in}^C + 2S_{in}^a}{(S_{in}^C + S_{in}^a+S_{out}^c + S_{out}^a)} - \frac{S_{in}^C}{S_{in}^C + S_{out}^C}=\\
&= \frac{2S_{in}^a \cdot S_{out}^C - S_{in}^C \cdot S_{out}^a + S_{in}^C \cdot S_{in}^a} {(S_{in}^C + S_{in}^a+S_{out}^c + S_{out}^a) (S_{in}^C + S_{out}^C)}
\end{split}
\end{equation}
If $ \Delta T_C(A) > 0$, then $ 2S_{in}^a \cdot S_{out}^C - S_{in}^C \cdot S_{out}^a + S_{in}^C \cdot S_{in}^a > 0$ which is equivalent to $\frac{S_{out}^C}{S_{out}^C} - \frac{S_{out}^a - S_{in}^a}{2S_{in}^a}$. Then they define the tightness gain in the following~\cite{Huang11}.

{\bf Definition 4} \textit{(Tightness Gain) The tightness gain for the community C adopting a neighbor vertex {\large a} can be denoted as}
\begin{equation} 
\tau{_C}(A) = \frac{S_{out}^C }{S_{in}^C} - \frac{S_{out}^a - S_{in}^a}{2S_{in}^a}.
\end{equation}
It means that the ratio of external similarity to internal similarity of
community $C$ is greater than the ratio of external similarity
increment to internal similarity increment caused by adopting
vertex \textit{{\large a}}. Obviously, this case will result in the increase of the
tightness value of community $C$. Therefore, $\tau_{C} (a)$ can be utilized
as a criterion to determine whether the candidate vertex \textit{{\large a}} should
be included in the community $C$ or not.
In the following, they introduce an optional resolution parameter
$\beta$ to control the scale at which we want to observe the communities
in a network. 
 
{\bf Definition 5} \textit{(Tunable Tightness Gain) The tunable tightness gain for
the community $C$ merging a neighbor vertex {\large a} can be denoted as}
\begin{equation} 
\tau{_C}^{\alpha}(A) = \frac{S_{out}^C }{S_{in}^C} - \frac{\alpha S_{out}^a - S_{in}^a}{2S_{in}^a}.
\end{equation}

A parameter $\beta \in (0,\infty) $  is introduced as the coefficient of $S_{out}^a$
which can increase or decrease the proportion of the external
similarity of the candidate vertex \textit{{\large a}}. Here, the criterion for
accepting a vertex \textit{{\large a}} is changed to $ \tau{_C}^{\alpha}(A) >0$. For $\alpha = 1$ , the criteria
is moderate and can be used in most normal cases. In~\cite{Huang11} the authors shows different scenarios for different values of the free parameter $\beta$:
setting $\beta \in (0,1)$, the value of $S_{out}^a$ is reduced by this coefficient which
increases the chance of a candidate vertex \textit{{\large a}} joining $C$ and bigger
communities will be formed compared to the normal case with
$\beta = 1 $. On the contrary, it will result in the formation of smaller
communities in a network when we set $\beta >1 $.

\end{document}